\documentclass[11pt]{article}
\usepackage{a4wide}

\usepackage[dvipdfmx,hiresbb]{graphicx}
\usepackage{mediabb}

\usepackage{color}

\usepackage{amsmath}
\usepackage{amssymb}
\usepackage{latexsym}
\usepackage{braket}
\usepackage{float}
\usepackage{comment}

\newcommand{\mc}{\mathcal}

\DeclareMathOperator{\tr}{tr}
\DeclareMathOperator{\Tr}{Tr}
\DeclareMathOperator{\STr}{STr}
\DeclareMathOperator{\str}{str}
\DeclareMathOperator{\sdet}{sdet}
\DeclareMathOperator{\SDet}{SDet}
\DeclareMathOperator{\Det}{Det}

\DeclareMathOperator{\re}{Re}
\DeclareMathOperator{\im}{Im}

\newcommand{\T}{\text{T}}
\newcommand{\wt}{\widetilde}

\newcommand{\BB}{\text{BB}}
\newcommand{\BF}{\text{BF}}
\newcommand{\FB}{\text{FB}}
\newcommand{\FF}{\text{FF}}

\newcommand{\mO}{\mathcal{O}}

\newcommand{\ol}{\ensuremath{\overline}}

\newcommand{\al}[1]{\begin{align}#1\end{align}}
\newcommand{\als}[1]{\begin{align*}#1\end{align*}}

\newcommand{\bp}{\begin{pmatrix}}
\newcommand{\ep}{\end{pmatrix}}
\newcommand{\nn}{\nonumber\\}
\newcommand{\paren}[1]{\left(#1\right)}
\newcommand{\sqbr}[1]{\left[#1\right]}

\newcommand{\p}{\partial}

\newcommand{\GeV}{\,\text{GeV}}

\newcommand{\cred}{}
\newcommand{\cblue}{}

\newcommand{\df}{\text{d}}
\newcommand{\ab}[1]{\left|#1\right|}
\newcommand{\bs}[1]{\boldsymbol}

\newcommand{\pmat}[1]{\begin{pmatrix}#1\end{pmatrix}}
\newcommand{\bmat}[1]{\begin{bmatrix}#1\end{bmatrix}}
\newcommand{\fn}[1]{\!\left(#1\right)}

\newcommand{\Slash}[1]{{\ooalign{\hfil/\hfil\crcr$#1$}}} 

\newcommand{\wh}{\widehat}

\newcommand{\ft}{\tilde} 
\newcommand{\tildeForLaplace}{\tilde}
\newcommand{\ch}{\hat} 

\newcommand{\ola}{\overleftarrow}

\usepackage{etoolbox}
\let\bbordermatrix\bordermatrix
\patchcmd{\bbordermatrix}{8.75}{4.75}{}{}
\patchcmd{\bbordermatrix}{\left(}{\left[}{}{}
\patchcmd{\bbordermatrix}{\right)}{\right]}{}{}

\begin{document}
\title{
\vbox{
\baselineskip 14pt
\hfill \hbox{\normalsize OU-HET/873-2015}\\
\hfill \hbox{\normalsize KANAZAWA-15-15}
}		
\vskip 1cm
Non-minimal coupling in Higgs-Yukawa model\\
with asymptotically safe gravity
\bigskip
}
\author{
	Kin-ya Oda\thanks{E-mail: \tt odakin@phys.sci.osaka-u.ac.jp}
	\ and
	Masatoshi Yamada\thanks{E-mail: \tt masay@hep.s.kanazawa-u.ac.jp}
	\\
	\\
	$^{*}$\it\normalsize Department of Physics, Osaka University, Osaka 560-0043, Japan\\
	$^\dagger$\it\normalsize Institute for Theoretical Physics, Kanazawa University, Kanazawa 920-1192, Japan\\
	\bigskip\\}
\date{}
\maketitle
\begin{abstract}\noindent
We study the fixed point structure of the Higgs-Yukawa model, with its scalar being non-minimally coupled to the asymptotically safe gravity, using the functional renormalization group.
We have obtained the renormalization group equations for the cosmological and Newton constants, the scalar mass-squared and quartic coupling constant, and the Yukawa and non-minimal coupling constants, taking into account all the scalar, fermion, and graviton loops. We find that switching on the fermionic quantum fluctuations makes the non-minimal coupling constant irrelevant around the Gaussian-matter fixed point with the asymptotically safe gravity.
\end{abstract}

\newpage
\section{Introduction}
Construction of quantum gravity is one of the most important, and challenging, subjects in physics. The general relativity is derived from the Einstein-Hilbert action 
\al{
S_{\rm EH}=\int \df^Dx\,\sqrt{g} \left( - {\frac{1}{16\pi G}R}  +\lambda \right),
	\label{EH action}
}
where $G$ and $\lambda$ are the Newton and cosmological constants and we work with the Euclidean action throughout this paper. The Einstein gravity~\eqref{EH action} can accurately account for the macroscopic phenomena such as the perihelion precession of Mercury and the gravitational lensing. Therefore we believe that the action~\eqref{EH action} correctly describes the dynamics of gravity in the long range.
On the other hand, its quantization is quite difficult because of the non-renormalizability.
The asymptotically safe quantum gravity, suggested by Weinberg~\cite{Hawking:1979ig}, is one of the possible candidates of quantum gravity.

\begin{figure}
\begin{center}
\includegraphics[width=0.53\textwidth]{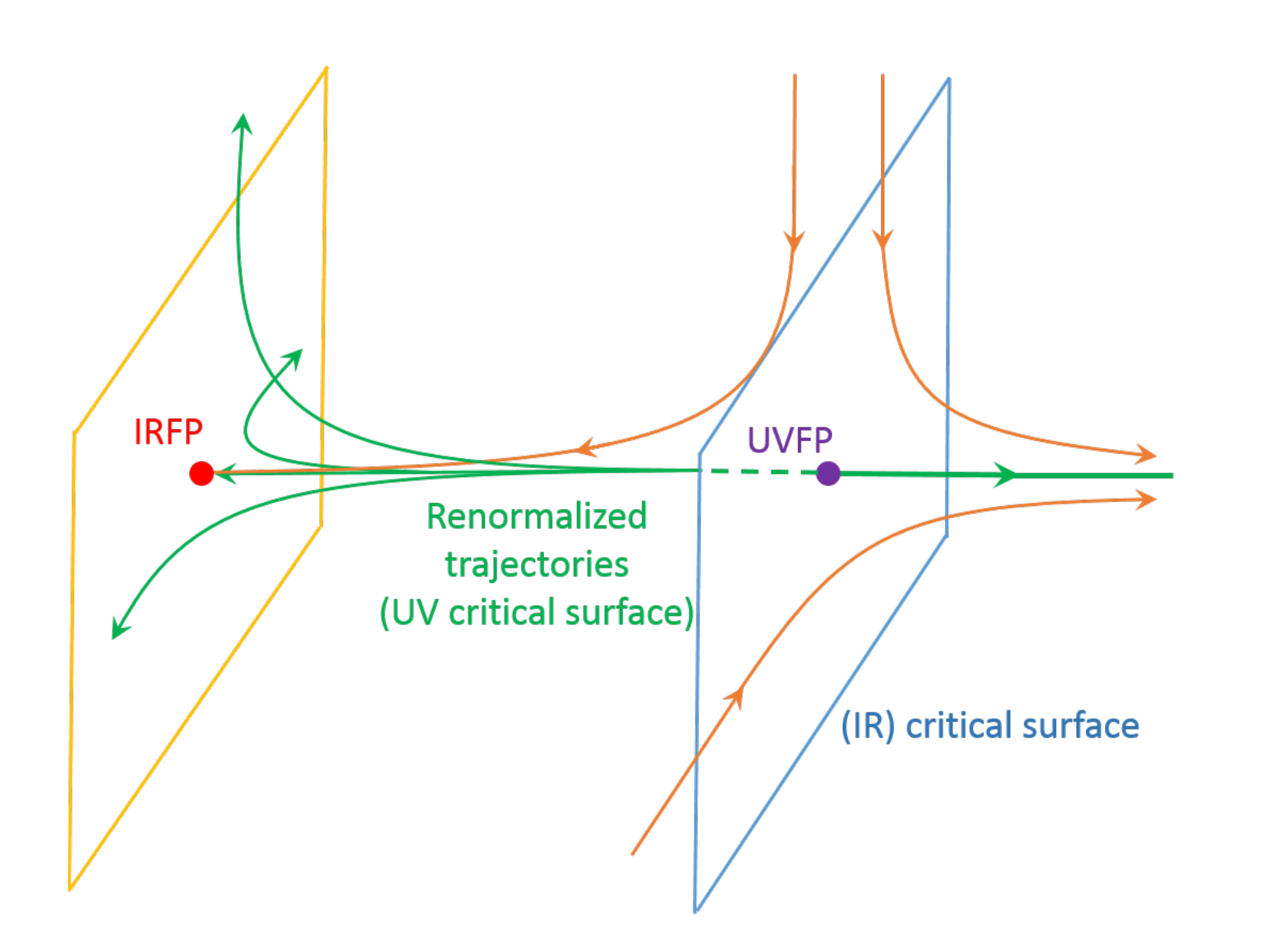}
\end{center}
\caption{Schematic figure for the RG flow in the theory space. 
The arrows indicate the direction from UV to IR.
The left (red) and right (purple) points, labelled IRFP and UVFP, are the IR and UV fixed points, respectively.
The UV critical surface (green) is the finite-dimensional subspace spanned by the renormalized trajectories that are flowing out of the UV fixed point.
Under the asymptotic safety, our low energy effective theory is one of the points on the renormalized trajectory.
The right (blue) surface is the IR critical surface, which is generally infinite dimensional; see footnote~\ref{IR critical surface}.
The other (orange) generic flows cannot be used to construct an asymptotically safe theory.
The left (yellow) surface is a finite-dimensional subspace spanned by the relevant directions around the IR fixed point.
}
\label{rgflow}
\end{figure}

It is essential for the scenario of asymptotic safety that there exists a non-trivial ultraviolet (UV) fixed point.\footnote{
In general we call a fixed point ``UV'' if it has a relevant direction.
(Throughout this paper, we call an operator ``relevant'' if its coupling constant departs from the fixed point in the flow from UV to IR.)
Sometimes ``IR fixed point'' is defined by the condition that all the directions become either irrelevant or marginal around it. Here instead we call a fixed point ``IR'', even when there exists a relevant direction around it, if the RG flow from the UV fixed point is attracted toward it, as is the case for the Wilson-Fisher fixed point. See Fig.~\ref{rgflow}.
}
Around the UV fixed point, two hypersurfaces are defined: the UV and infrared (IR) critical surfaces.\footnote{\label{IR critical surface}
Usually, ``critical surface'' refers to what is called the ``IR critical surface'' here. In the literature on the asymptotic safety, the wording ``UV critical surface'' is used frequently, and we put ``IR'' on what is usually called ``critical surface'', in order to distinguish it from the other.
}
The UV critical surface consists of the renormalized trajectories that are flowing out of the UV fixed point and is in general finite dimensional, whereas the IR critical surface is its orthogonal complement and is infinite dimensional in general. See Fig.~\ref{rgflow}. 

The renormalization group (RG) flow of the renormalized trajectory on the UV critical surface takes infinite steps of renormalization transformations near the UV fixed point. If the IR physics is realized as a point on the UV critical surface, then the continuum limit $\Lambda\to \infty$ can be taken, and the theory is free from UV divergences.
Furthermore, when the dimension of the UV critical surface is finite, the theory is non-perturbatively renormalizable even if it is non-renormalizable in perturbation theory; see e.g.\ Refs.~\cite{Percacci:2007sz,Percacci:2011fr,Braun:2010tt,Nagy:2012ef,Ambjorn:2012jv}.
The idea of the asymptotic safety has been applied not only to gravity but also to the extra-dimensional model~\cite{Kubo:1999ua,Kubo:2000hy,Kubo:2001tr,Gies:2003ic} and to the Higgs-Yukawa model in flat spacetime~\cite{Gies:2009hq,Gies:2009sv,Scherer:2009wu,Litim:2014uca,Gies:2013pma}.
The quantum Einstein gravity theory is asymptotically safe if there exists the UV critical surface including the Newton constant.
We will further review the concept of the asymptotic safety in section~\ref{asreview}.

In earlier study the exsistence of UV fixed point of the Newton constant $G$ has been studied by an $\epsilon$-expansion in $2+\epsilon$ dimensions~\cite{Hawking:1979ig,Kawai:1989yh}. 
The fixed point of the dimensionless rescaled Newton constant $\tilde G:=G\Lambda^\epsilon$ is found as $\tilde G^*= 3\epsilon/38$ when ignoring the cosmological constant $\lambda$. The dimensionful Newton constant $G$ vanishes asymptotically around the fixed point if $\epsilon >0$, that is, $G\simeq\tilde G^*/\Lambda^\epsilon \to 0$ for $\Lambda\to \infty$.
Then the theory is asymptotically free.

The $\epsilon$ expansion method has difficulties in applying to arbitrary space-time dimensions and in analyzing the theory in detail. The functional renormalization group (FRG)~\cite{Aoki:2000wm,Bagnuls:2000ae,Berges:2000ew,Pawlowski:2005xe,Gies:2006wv,Rosten:2010vm} is useful for such purposes.
After its pioneering application to the quantum Einstein gravity given by Reuter~\cite{Reuter:1996cp}, the UV fixed point and the RG flow structure of the Einstein gravity have been investigated in Refs.~\cite{Dou:1997fg,Souma:1999at,Reuter:2001ag,Lauscher:2001rz,Lauscher:2001cq,Lauscher:2001ya,Litim:2003vp,Reuter:2005bb,Fischer:2006fz,Reuter:2007rv,Reuter:2008wj,Reuter:2008qx,Benedetti:2009gn,Groh:2010ta,Eichhorn:2010tb,Manrique:2010am,Manrique:2011jc,Harst:2012ni,Donkin:2012ud,Dona:2012am,Christiansen:2012rx,Christiansen:2014raa,Christiansen:2015rva,Gies:2015tca}, and its extended models with matters are studied in Refs.~\cite{Percacci:2002ie,Percacci:2003jz,Bonanno:2004sy,Narain:2009fy,Narain:2009gb,Zanusso:2009bs,Daum:2009dn,Eichhorn:2009ah,Daum:2010bc,Manrique:2010mq,Vacca:2010mj,Folkerts:2011jz,Harst:2011zx,Eichhorn:2011pc,Eichhorn:2012va,Hindmarsh:2012rc,Dona:2013qba}; see also Refs.~\cite{Niedermaier:2006wt,Niedermaier:2006ns,Litim:2008tt,Litim:2011cp,Reuter:2012id,Codello:2008vh} for reviews.\footnote{
In Refs.~\cite{Falkenberg:1996bq,Souma:2000vs,Gies:2015tca}, the issue of gauge dependence has been discussed.
}
The existence of the UV fixed point and the stability of the dimension of the UV critical surface when extending the theory space have been studied in Refs.~\cite{Lauscher:2002sq,Codello:2006in,Codello:2007bd,Machado:2007ea,Benedetti:2009rx,Benedetti:2009iq,Bonanno:2010bt,Niedermaier:2010zz,Dietz:2012ic,Ohta:2012vb,Benedetti:2013nya,Falls:2013bv,Ohta:2013uca,Dietz:2013sba,Falls:2014tra,Ohta:2015efa,Labus:2015ska,Eichhorn:2015bna}.
For example in Ref.~\cite{Falls:2014tra}, the $f(R)$ gravity that has powers of the Ricci scalar $R$ up to order $R^{34}$ is studied, and it has been shown that the number of dimensions of the UV critical surface is stable to be three, namely, the relevant operators are $\lambda$, $R$ and $R^2$.
Furthermore, the Higgs mass was predicted to be $\simeq 126\GeV$ before the Higgs discovery by requiring the Higgs quartic coupling to vanish around the Planck scale in the context of the asymptotically safe gravity~\cite{Shaposhnikov:2009pv}. 
These results encourage the asymptotic safety scenario for the quantum gravity.

The purpose of this paper is to contribute to the investigation whether the asymptotically safe gravity can have a large non-minimal coupling $\xi$ between $R$ and a scalar field in the IR limit.
Such a large non-minimal coupling plays a crucial role in the Higgs inflation scenario~\cite{Bezrukov:2007ep,Salvio:2013rja,Hamada:2013mya,Cook:2014dga,Hamada:2014iga,Bezrukov:2014bra,Hamada:2014xka,Hamada:2014wna,Hamada:2014raa,Ibanez:2014swa,Bezrukov:2014ipa,Salvio:2015kka}; see Ref.~\cite{Nielsen:2015una} for a phenomenological study in a concrete model of the Higgs inflation under the asymptotically safe gravity.\footnote{
In Refs.~\cite{Copeland:2013vva,Bonanno:2015fga}, the asymptotically safe gravity has been applied to the Starobinsky $R^2$ inflation model.
In Ref.~\cite{Xianyu:2014eba}, it has been claimed that the Higgs potential becomes flat above a certain transition scale under the asymptotic safety.
See Refs.~\cite{Weinberg:2009wa,Tye:2010an,Bonanno:2010mk,Fang:2012ca} for attempts of the so-called asymptotically safe inflation, and also Refs.~\cite{Cai:2012qi,Cai:2013caa}.
}
In the first attempts of Higgs inflation~\cite{Bezrukov:2007ep,Salvio:2013rja}, it was necessary to have an extremely large value of $\xi$ of order $10^4$--$10^5$ to account for the cosmological data. Later it has been pointed out~\cite{Hamada:2014iga,Bezrukov:2014bra} that it is possible to have a successful Higgs inflation with smaller $\xi\sim10$, given the criticality of the Higgs potential, i.e., the fact that both the Higgs quartic coupling and its beta function can vanish at around the Planck scale $\sim \paren{32\pi G}^{-1/2}\simeq10^{18}\,\text{GeV}$; see also Ref.~\cite{Cook:2014dga}. For the criticality of the Higgs potential in the Standard Model, it is essential that the Higgs field has the large Yukawa coupling to the top quark. Therefore, it is important to understand the asymptotic safety in a Higgs-Yukawa system which is non-minimally coupled to the gravity.

In Refs.~\cite{Percacci:2003jz,Narain:2009fy,Narain:2009gb,Henz:2013oxa,Shapiro:2015ova}, the authors have analyzed a simplified scalar-gravity system without fermions, taking into account the non-minimal coupling $\xi$ between a neutral scalar and the Ricci scalar, under the local potential approximation (LPA) in FRG: The non-minimal coupling is shown to be  \emph{relevant} around the UV fixed point. In Ref.~\cite{Zanusso:2009bs}, the authors have analyzed a simplified Higgs-Yukawa system with the same neutral scalar and an additional fermion and without $\xi$, in the flat spacetime. 
We combine these two approaches and analyze the running of $\xi$ and the Yukawa coupling under the influence of the fermionic quantum fluctuations, in the simplified Higgs-Yukawa system that is non-minimally coupled to gravity.
We find that $\xi$ becomes \emph{irrelevant} by inclusion of the fermions.

This paper is organized as follow: 
We briefly review the concept of asymptotic safety in the next section.
In section~\ref{models}, we introduce the Higgs-Yukawa model which is non-minimally coupled to gravity.
In Sec.~\ref{RG equations of system}, we show explicitly the RG equations of the model.
In section~\ref{numericalan}, we present the methods and results of the numerical analysis.
In Sec.~\ref{summary}, we give summary and discussions.
In appendix~\ref{FRGnote}, we briefly sketch how the Wetterich equation is derived from the cutoff dependence of the effective action equipped with the cutoff function.
In appendix~\ref{supermatrix}, we list the formulae for supermatrix.
In appendix~\ref{frgtreatment}, we rewrite the Wetterich equation into suitable form to be used in our application, using the supermatrix formula.
In appendix~\ref{hkeapp}, we review the heat kernel expansion techniques which are used to sum up the eigenvalues of the differential operators.
In appendix~\ref{explicitcal}, we show the explicit derivations of the beta functions in our system.

\section{Asymptotic safety}\label{asreview}
In this section we explain the basic idea of the asymptotic safety.
We start from a system described by an effective action
\al{\label{effectiveactiongeneral}
\Gamma_\Lambda = \int \df ^D x \displaystyle \sum_i^\infty \frac{g_{i, \Lambda}}{\Lambda^{D_{\mathcal O_i}-D}} \mathcal O_i,
}  
where $g_{i,\Lambda}$ are the dimensionless coupling constants, $\mathcal O_i$ are operator bases, and $D_{\mathcal O_i}$ is the dimension of $\mathcal O_i$.
Let us write the RGE for the coupling constant $g_{i,\Lambda}$
\al{\label{rgequations}
-\Lambda \frac{ \p g_{i,\Lambda}}{\p \Lambda} 
	=	\beta_i\fn{g_{\Lambda}},
}
where $\beta_i\fn{g_{\Lambda}}$ is the abbreviation for 
$\beta_i\fn{g_{1,\Lambda},\,g_{2,\Lambda},\,\dots}$.
The fixed point $g^*$ is given by the solution to the vanishing beta functions:
\al{
\beta_i\fn{g^*}=0.
	\label{fixed point equation}
}
In many cases, there exists the trivial (Gaussian) fixed point: $g_i^*=0$ for all $i$.

Here we consider the case that the coupled equation~\eqref{fixed point equation} has a non-trivial fixed point with $g_{j}^*\neq0$ for some $j$ and that this is the UV fixed point, namely, there exists a relevant direction flowing out of this point.
The resultant RG flows, given by Eq.~\eqref{rgequations}, are as schematically shown in Fig.~\ref{rgflow}.
As said in Introduction, the IR critical surface separates the theory space (the space of coupling constants) into two phases, and is spanned by an infinite number of irrelevant operators.
On the other hand, the UV critical surface (the green renormalized trajectories in Fig.~\ref{rgflow}) controls the IR physics. That is, an arbitrary RG flow from the neighborhood of UV fixed point approaches  this hypersurface at IR scales.
In other words, when we fix the physics at IR scales and take the continuum limit $\Lambda \to \infty$, the theory on the renormalized trajectory approaches the UV non-trivial fixed point and dose not diverge from it.

To conclude, such an RG flow can be a candidate of the UV complete theory.
Furthermore, if the dimension of the UV critical surface is finite, the theory is renormalizable:
The finite number of parameters spanning the UV critical surface determine all other parameters.\footnote{The renormalizability in low energy region is guaranteed by an existence of  the stable hypersurface with finite dimention. It is known as the Polchinski theorem; see~\cite{Polchinski:1983gv,Hurd:1989up,Keller:1990ej} for the scalar theory and \cite{Keller:1995qn} for QED.}
Note that a perturbative expansion can be done only at the vicinity of the trivial fixed point $g_{i,\Lambda}^*=0$ and that we need a non-perturbative method to analyze the whole structure the RG flows to find out the fixed points.

To see the renormalizability of the theory, we evaluate the critical exponents $\theta_i$ of the coupling constants by linearizing the RGEs~\eqref{rgequations} around the fixed point $g_{i,\Lambda}^*$:
\al{\label{fixedrg}
g_{i,\Lambda} = g_{i,\Lambda}^* + \displaystyle \sum_{j=0}^\infty \zeta_{ij} \left( \frac{\Lambda_0}{\Lambda} \right) ^{\theta_j},
}
where $\Lambda_0$ is a UV cutoff scale.\footnote{
An explicit derivation is shown in Sec.~\ref{numericalan}.
}
The RG flow going away from the fixed point has a critical exponent with positive real part,\footnote{
The imaginary part of the critical exponent corresponds to the mixing with other couplings when flowing out of the UV fixed point.
}
and is on the UV critical surface. 
Therefore, we can examine whether the UV critical surface is finite dimensional or not, by investigating the number of positive critical exponents.
We will investigate the fixed point structure and the critical exponent of our theory in Sec.~\ref{numericalan}.

Let us illustrate the situation with the case where the UV critical surface is spanned by a single operator $\mathcal O_1$.
The dimensionful parameter $G_{1,\Lambda}$ reads
\al{
g_{1,\Lambda}^* = \Lambda ^{D_{\mathcal O_1}-D} G_{1,\Lambda}.
}
When $D_{\mathcal O_1}-D >0$, the dimensionful parameter $G_{1,\Lambda}$ at vicinity of the UV fixed point goes to zero in the UV limit $\Lambda \to \infty$. That is, the theory becomes asymptotically free.
When $D_{\mathcal O_1} =D$, the coupling constant $G_{1,\Lambda}$ is dimensionless and the theory becomes asymptotically non-free.\footnote{
If the fixed point $g_{1,\Lambda}^*$ is a trivial UV fixed point $g_{1,\Lambda}^*=0$ with $D_{\mathcal O_1} =D$, the coupling constant $G_{1,\Lambda}$ in the UV limit, and the theory becomes asymptotically free. This is the case for the quantum chromodynamics.
}
To summarize, the asymptotic safety is a generalization of the asymptotic freedom.
We have $\zeta_{1j} = g_{1,\Lambda} \delta_{1j}$ with a positive critical exponent $\rm Re(\theta_1)>0$, while the others are negative, and $g_{1,\Lambda}$ becomes a single physical free parameter of the theory.

We comment on the dimension of an operator and its coupling constant.
The beta function of $g_{1,\Lambda}$ is typically given as\footnote{
The FRG is one-loop exact and the term of $\mathcal O\fn{g_{1,\Lambda}^3}$ does not appear in general.
}
\al{
\beta_{g_{1,\Lambda}}= -\paren{D_{\mathcal O_1}-D} g_{1,\Lambda} + Lg_{1,\Lambda}^2,
	\label{beta of g1}
}
where $L$ is a loop factor and we have ignored the contributions from other couplings.
Note that the anomalous dimension from the field renormalization is ignored throughout this paper as we are taking the LPA.
Around the trivial fixed point $g_{1,\Lambda}^*=0$, the first term (the so-called canonical scaling term) in Eq.~\eqref{beta of g1} becomes dominant. The coefficient $-\paren{D_{\mathcal O_1}-D}$  is the dimension of $g_{1,\Lambda}$, and the dimension of the operator is simply $D_{\mathcal O_1}$.
On the other hand, at the non-trivial fixed point, 
we get $g_{1,\Lambda}^*=\paren{D_{\mathcal O_1}-D}/L$, and the beta function is rewritten as
\al{
\beta_{g_{1,\Lambda}}= \paren{D_{\mathcal O_1}-D}\paren{g_{1,\Lambda}-g_{1,\Lambda}^*} + L\paren{g_{1,\Lambda}-g_{1,\Lambda}^*}^2.
	\label{beta at fixed point}
}
The dimension of operator is effectively \cblue{changed to $D-\paren{D_{\mathcal O_1}-D}=2D-D_{\mathcal O_1}$}.

Let us consider the case of the gravity in $D=4$.
The operator ${\mathcal O_1}=R$ and its coupling constant $g_{1,\Lambda}=1/16\pi G$ have the canonical dimension $D_{\mathcal O_1}=2$ and $-\paren{D_{\mathcal O_1}-D}=2$, respectively. 
\cblue{At the non-trivial fixed point, the effective dimension of the operator becomes \mbox{$2D-D_{\mathcal O_1}=6$}, and that of the gravitational coupling constant $g_{1,\Lambda}$ has been changed from the canonical dimension $-\paren{D_{\mathcal O_1}-D}=2$ to the value \mbox{$D_{\mathcal O_1}-D=-2$}. The critical exponent $D_{\mathcal O_1}-D$ in Eq.~\eqref{beta at fixed point} is physically the effective dimension of the coupling around the non-trivial fixed point.}

There are attempts to read off the number of effective degrees of freedom (namely the spectral dimension) from the RG flows of the theory, in order to test whether the asymptotically safe gravity can be achieved or not:
Such attempts have been made in Refs.~\cite{Lauscher:2005qz,Reuter:2011ah,Reuter:2012xf,Reuter:2012id,Rechenberger:2012pm} using the FRG and
in Refs.~\cite{Ambjorn:2005db,Laiho:2011ya} using the lattice simulation; see also Refs.~\cite{Egawa:1996ff,Egawa:1996fu,Horata:2002uf,Horata:2003hm,Egawa:2003gk} for related studies.

\section{Non-minimal Higgs-Yukawa model}\label{models}

\subsection{The model}

As a toy model for the Higgs inflation scenario under the asymptotically safe gravity, we study a Higgs-Yukawa model with a real scalar field $\wh\phi$ and with $N_\text{f}$-flavors of Dirac fermions~$\wh\psi$, where its flavor index is suppressed. We write the metric $\wh g_{\mu\nu}$ and the volume element $\sqrt{\wh g}$.
We decompose the integration variables $\wh g_{\mu\nu}$, $\wh\phi$ and $\wh\psi$ in the functional integral over all field configurations according to
\al{
\wh g_{\mu\nu}
	&=	g_{\mu\nu}+h_{\mu\nu},\nn
\wh\phi
	&=	\phi+\varphi,\nn
\wh\psi
	&=	\psi+\chi,
}
where $g_{\mu\nu}$, $\phi$ and $\psi$ are fixed background fields so that the integration over $\wh g_{\mu\nu}$, $\wh\phi$ and $\wh\psi$ may be replaced by an integration over $h_{\mu\nu}$, $\varphi$ and $\chi$, respectively.

We write the truncated effective action in Euclidean space:
\al{
\Gamma_\Lambda[g_{\mu\nu},\phi,\psi;\,h_{\mu\nu},\varphi,\chi]
	&=	\int\df^4x \sqrt{\wh g}\Bigg\{ V_\Lambda\fn{\wh\phi ^2}-F_\Lambda \fn{\wh\phi ^2}\wh R
		+\frac{1}{2}\wh g^{\mu \nu}\,\partial _\mu{\wh \phi}\,\partial _{\nu} \wh\phi 
		+\ol{\wh\psi}\wh{\Slash D}\wh\psi	
		+y_\Lambda \wh\phi \ol{\wh\psi}\wh\psi
		\Bigg\}\nn
	&\quad
		+S_{\rm GF}	+S_{\rm gh},
		\label{effectiveaction}
}
where $\p_\mu$ and $\wh{\Slash D}$ are the general covariant derivatives on the scalar fields and on spinor fields that includes the spin connection;
$S_\text{GF}$ and $S_\text{gh}$ are the gauge fixing and ghost terms, respectively, shown below; and
the widehat symbol $\wh{\mbox{ }}$ denotes that the corresponding quantity is made of the  metric $\wh g_{\mu\nu}$ and veirbein $\wh e_\mu^a$. 
We have imposed the $Z_2$ symmetry: $\wh\phi\to-\wh\phi$ and $\wh\psi\to\gamma_5\wh\psi$.\footnote{
A background $\phi\neq0$ breaks this $Z_2$ symmetry. In this paper, we restrict our attention to the case $\phi=0$.
}

We expand the scalar potential and the non-minimal coupling of $\wh\phi$ to the gravity:
\al{
V_\Lambda\fn{\wh\phi^2}
	&=	\ch \lambda_0\fn{\Lambda}
		+\ch \lambda_2\fn{\Lambda}\,\wh\phi^2
		+\ch \lambda_4\fn{\Lambda}\,\wh\phi^4+\cdots,	\\
F_\Lambda\fn{\wh\phi^2}
	&=	\ch \xi_0\fn{\Lambda}+\ch \xi_2\fn{\Lambda}\,\wh\phi^2+\ch \xi_4\fn{\Lambda}\,\wh\phi^4+\cdots.
}
In more conventional language, $\ch\lambda_0$ is the cosmological constant; $\ch\lambda_2=m^2/2$  gives the mass parameter of the scalar field; and $\ch\xi_0=1/16\pi G$ the Newton constant.
The non-minimal coupling $\ch\xi_2$ plays a crucial role in the Higgs inflation scenario~\cite{Bezrukov:2007ep}; see also Ref.~\cite{Nielsen:2015una}.

We employ the following gauge-fixing and ghost actions for the diffeomorphisms~\cite{Reuter:1996cp,Percacci:2003jz,Narain:2009fy,Narain:2009gb} 
\begin{align}
S_{\rm GF}	&=	\frac{1}{2\alpha}\int \df^Dx\sqrt{ g}\,
					F\fn{\phi ^2}{ g}^{\mu \nu}\Sigma_\mu\Sigma_{\nu},
					\label{gaugefixedaction} \\
S_{\rm gh}	&=	-\int\df^Dx\sqrt{ g}\,\bar C_\mu\left[ g^{\mu\rho}{ \p}^2+\frac{1-\beta}{2}{ \p }^\mu{ \p}^{\rho}		+{ R}^{\mu\rho}\right] C_{\rho}, \label{ghostaction}
\end{align}
where $C_\mu$ and $\bar C_\mu$ are the ghost and anti-ghost fields for the diffeomorphisms, respectively; $\alpha$ and $\beta$ are gauge parameters; and
\al{
\Sigma_\mu		
	&:= \p^\nu h_{\nu \mu}-\frac{\beta +1}{D}{ \p}_\mu h,
}
with $h$ being the trace part of the fluctuation $g^{\mu\nu}h_{\mu\nu}$.
Throughout this paper, the expression without the widehat symbol $\wh{\mbox{ }}$  indicates that the indices are raised and lowered by the background metric $g_{\mu\nu}$, and expressions such as $R$ and $\Slash D$ are written in terms of $g_{\mu\nu}$ and the background vierbein $e_\mu^a$.

\subsection{Two-point functions}\label{modelandmanipulations}

We collectively write the background fields $\Phi:=\paren{g_{\mu\nu},\phi,\psi}$ and the fluctuations $\Upsilon:=\paren{h_{\mu\nu},\varphi,\chi,C_\mu,\bar C_\mu}$.\footnote{
The ghost $C_\mu$ and anti-ghost $\bar C_\mu$ are treated as fluctuations only.
}
The effective action is written as $\Gamma_\Lambda[\Phi;\Upsilon]$, which is expanded as
\al{
\Gamma_\Lambda[\Phi;\Upsilon]=\Gamma_\Lambda[\Phi] + \Gamma_\Lambda^{(1)}[\Phi;\Upsilon]+
 \Gamma_\Lambda^{(2)}[\Phi;\Upsilon]+\mO\fn{\Upsilon^3},
}
where $\Gamma_\Lambda^{(n)}[\Phi;\Upsilon]$ contains the terms of order $\Upsilon^n$.

To derive the beta functions for the Higgs-Yukawa model, we need to evaluate the $\Gamma_\Lambda^{(2)}$ terms.
After some computations, we obtain 
\al{
\Gamma_\Lambda^{(2)}[\Phi;\Upsilon]
	&=\frac{1}{2}\int \df^4x \sqrt{ g}\bigg[ 
		-\frac{1}{2}F\fn{\phi^2} h^{\mu\nu} \p^2 h_{\mu\nu}
		+ \frac{1}{2}F\fn{\phi^2} h  \p^2 h 
		-F\fn{ \phi^2} h \p_\mu \p_\nu h^{\mu\nu}
		+F\fn{ \phi^2} h^{\mu\nu} \p_\mu\p_\rho h^\rho_{~\nu} \nn
	&\phantom{=\frac{1}{2}\int \df^4x \sqrt{ g}\bigg[}
 		+\bigg( \frac{1}{4}h^2 -\frac{1}{2}h_{\mu\nu}h^{\mu\nu} \bigg) \left( V\fn{ \phi^2} +y \phi \ol\psi \psi -F\fn{ \phi^2}  R \right) \nn
	&\phantom{=\frac{1}{2}\int \df^4x \sqrt{ g}\bigg[}
		+F\fn{ \phi^2} h h^{\mu\nu } R_{\mu\nu}
		-F\fn{ \phi^2}h_{\rho}^{~\nu}h^{\mu\rho} R_{\mu\nu}
		-F\fn{\phi^2}h^{\mu\nu} R_{\rho \mu \sigma \nu}h^{\rho \sigma} \nn
	&\phantom{=\frac{1}{2}\int \df^4x \sqrt{ g}\bigg[}
		-\frac{1}{16}h^{~\mu}_{\rho}\p_\nu h_{\sigma \mu}\ol\psi\gamma^\nu[\gamma^\rho,\gamma^\sigma]\psi
	\bigg] \nn
&\quad +\int \df^4x \sqrt{ g}\, \varphi\bigg[ 
		-2\phi F'\fn{\phi^2} \left\{  \p_\mu  \p_\nu  -\p^2  g_{\mu\nu} \right\} h^{\mu\nu} \nn
&\phantom{\quad +\int \df^4x \sqrt{ g}\, \varphi\bigg[ }
	+h \left\{ \phi V'\fn{\phi^2} +\frac{1}{2}y  \ol\psi \psi -\phi F'\fn{\phi^2} R \right\}
	+h^{\mu\nu}\left\{ 2\phi F'\fn{\phi^2} +  R_{\mu\nu}\right\} 
	\bigg]\nn
&\quad + \int \df^4 x\sqrt{ g}\,
	h\bigg[ 
	\frac{1}{2}y\phi \paren{ \ol\psi \chi + \ol\chi \psi}
	\bigg]\nn
&\quad +\frac{1}{2}\int \df^4x \sqrt{ g} \, \varphi \bigg[ \left\{ - \p^2 +2V' \fn{\phi^2} +4\phi^2 V''\fn{\phi^2} \right\}  - R \left\{ 2F'\fn{\phi^2} +4\phi^2 F''\fn{\phi^2} \right\}
\bigg] \varphi \nn
&\quad +\int \df^4x \sqrt{ g}\bigg[
\frac{1}{4} \left(
-\p_\mu h + \p _\nu h^{\nu}_{~\mu}
\right)
\paren{ \ol\psi \gamma^\mu \chi -  \ol\chi \gamma^\mu \psi}
\bigg]\nn
&\quad+ \int \df^4x \sqrt{ g}\, \varphi\bigg[ y \paren{ \ol\psi \chi +  \ol\chi \psi } \bigg] + \int \df^4x\sqrt{ g} \,  \ol\chi \bigg[ \Slash{  \p} + y\phi\bigg]  \chi +S_{\rm GF} + S_{\rm gh},
	\label{2daction}
}
where $S_{\rm GF}$ and $S_{\rm gh}$ are given in Eqs.~(\ref{gaugefixedaction}) and (\ref{ghostaction}), respectively,\footnote{
They are already bilinear terms of the fluctuations.
}
and the prime symbol $'$ denotes a derivative with respect to $\phi^2$ so that
\al{
V_\phi
	&=	2\phi V',	&
V_{\phi\phi}
	&=	2V'+4\phi^2V'',	\nn
F_\phi
	&=	2\phi F',	&
F_{\phi\phi}
	&=	2F'+4\phi^2F''.
}
Inserting (\ref{metricdecomposition}) into $h_{\mu\nu}$ of (\ref{2daction}), we get the two-point functions for each field. We write down their explicit forms below.

\subsection{York decomposition}
We decompose the graviton fluctuation as~\cite{York:1973ia}
\al{\label{metricdecomposition}
h_{\mu\nu}	=	h_{\mu\nu}^\perp		+\p _\mu\ft\xi_{\nu}	+\p _{\nu}\ft\xi_\mu	
+\paren{\p_\mu\p_{\nu}	-\frac{1}{D} g_{\mu\nu}\p^2}\ft\sigma
 	+\frac{1}{D} g_{\mu\nu}h,
}
where $\p^2:= g^{\mu\nu}\p_\mu\p_\nu$;
 $h^\perp_{\mu\nu}$ is the transverse and traceless tensor field with spin 2;
 $\ft \xi_\mu$ is the transverse vector field with spin 1;
 and $\ft \sigma$ and $h:={ g}^{\mu\nu}h_{\mu\nu}$ are the scalar fields with spin 0.
These fields satisfy the following conditions: ${ g}^{\mu\nu}h^\perp_{\mu\nu}=0$, $\p^\nu h^\perp_{\mu\nu}=0$, and $\p^\mu \ft \xi_\mu=0$.

We decompose the ghosts into the transverse and scalar components:
\al{
C_\mu	
	&=	C_\mu^\perp			+\p_\mu\ft C,	\nn
\ft{\bar C}_\mu
	&=	\ft{\bar C}_\mu^\perp+\p_\mu{\bar C},
}
where $\ft C$, $\ft {\bar C}$ are spin-0 scalar fields and $C^\perp_\mu$, $\bar C_\mu ^\perp$ are spin-1 transverse vector fields that satisfy $\p^\mu C^\perp_\mu=\p^\mu {\bar C}^\perp_\mu=0$.

In order to absorb the Jacobean of the path integral measure from the above decompositions, we redefine several components of the fluctuations as follows:
\begin{align}	
\xi_\mu
	&=	\sqrt{-\p ^2-\frac{ R}{D}}\,\ft\xi_\mu,	&
{\sigma}	
	&=	\sqrt{-\p ^2-\frac{ R}{D-1}}\sqrt{-\p^2}\,\ft\sigma,	&
{C}	
	&=	\sqrt{-\p ^2}\,\ft C,&
{\bar C}	
	&=	\ft {\bar C}\sqrt{-\p ^2};&	
\end{align}
see e.g.\ Ref.~\cite{Codello:2008vh}.

To summarize, the degrees of freedom in our system are the spin two $h^\perp_{\mu\nu}$, the spin one $\xi_\mu$, $C^\perp_\mu$, $\bar C^\perp_\mu$ the spin half $\psi$, and the spin zero $\sigma$, $h$, $\phi$, $C$, $\bar C$.

\subsection{Explicit form of two-point functions}
For bosonic fields, we obtain
\al{
\Gamma_\BB
	&=	\bmat{
			\Gamma_{h^\perp_{\mu\nu} h^\perp_{\rho\sigma}}&0&0\\
			0	&	\Gamma_{\xi_\mu\xi_\nu}&0\\
			0	&	0	&	\Gamma_\text{SS}
			},
}
with
\al{
\Gamma_{h^\perp_{\mu\nu}h^\perp_{\rho\sigma}}
	&=	
			\paren{ g^{\mu\rho} g^{\nu\sigma}}_\text{sym}
				\sqbr{{F\over2}\paren{p^2+{2R\over3}}
			-{V+Y\over2}}+(\text{spin connection term}),
				\label{Gamma_hThT}\\
\Gamma_{\xi_\mu\xi_\nu}
	&=	 g^{\mu\nu}\sqbr{{F\over\alpha}\paren{p^2+{2\alpha-1\over4}R}-V-Y}+(\text{spin connection term}),
				\label{Gamma_xixi}\\
		\nn
\Gamma_\text{SS}
	&=	\bbordermatrix{
				&\sigma&h&\varphi\cr
			\sigma
				&	{\frac{3F}{16}\left( \frac{3-\alpha}{\alpha}p^2+\frac{\alpha-1}{\alpha}R\right)  -\frac{3\paren{V +Y}}{8}	}
 					&{  \frac{3F}{16}\frac{\beta-\alpha}{\alpha}\sqrt{p^2-\frac{R}{3}}\sqrt{p^2} 	}
					&{  -\frac{3F_\varphi}{4}\sqrt{p^2-\frac{R}{3}}\sqrt{p^2} }\cr
			h	&
{  \frac{3F}{16}\frac{\beta -\alpha}{\alpha}\sqrt{p^2-\frac{R}{3}} \sqrt{p^2}   }
 	 & { -\frac{F}{16}\frac{3\alpha -\beta ^2}{\alpha}p^2+\frac{V+Y}{8}}
	 	&{  -\frac{3F_\varphi}{4} \left( p^2+\frac{R}{3}\right) +{V_\varphi\over2}+{Y_{\cred\varphi}\over\cred2}} \cr
			\varphi&
			{  -\frac{3F_\varphi}{4}\sqrt{p^2-\frac{R}{3}} \sqrt{p^2}	}
 	& {  -\frac{3F_\varphi}{4} \left( p^2+\frac{R}{3}\right)+{V_\varphi\over2} +{Y_\varphi\over\cred2}}
		 &{ 	 p^2+V_{\varphi\varphi}-RF_{\varphi\varphi}  }
	},
}
where
$\paren{\cdots}_\text{sym}$ indicates that the indices inside parentheses are properly symmetrized;\footnote{
Explicitly,
$\paren{g^{\mu\rho}g^{\nu\sigma}}_\text{sym}
	=	{1\over4}\paren{
			g^{\mu\rho}g^{\nu\sigma}
			+g^{\nu\rho}g^{\mu\sigma}
			+g^{\nu\sigma}g^{\mu\rho}
			+g^{\mu\sigma}g^{\nu\rho}
			}$.
}
we write a minus of the d'Alembertian in the de Sitter space $p^2:=-\p^2$;
the over-left-arrow $\ola{\phantom\p}$ denotes that the differential operator acts on the left; and
$Y:=y\phi\ol\psi\psi$ and $Y_\phi:=y\ol\psi\psi$ are the Yukawa interaction and its derivative with respect to $\phi$, respectively.
The ``spin connection term'' in Eqs.~\eqref{Gamma_hThT} and \eqref{Gamma_xixi} is coming from the derivatives of the spin connection with respect to the metric, which only affect operators involving higher powers of $\psi$ and $\ol\psi$;
such operators are truncated in our effective action.

Other parts are given by
\al{
\Gamma_\BF
	&=	\bbordermatrix{
				&\chi&\smash{\ol\chi^\T}\cr
			h^\perp_{\mu\nu}
				&0&0\cr
			\xi_\mu
				&-\frac{1}{4}\sqrt{\ola p^2}\paren{\ol\psi\gamma^\mu}	&	-\frac{1}{4}\sqrt{\ola p^2}\paren{\gamma^\mu\psi}^\T\cr
			\sigma
				&-\frac{3}{16}\paren{\ola{\p_\mu}}\paren{\ol\psi\gamma^\mu} 	&	-\frac{3}{16}\paren{\ola{\p_\mu}}\paren{\gamma^\mu\psi}^\T\cr
			h	&\frac{y}{2}\phi {\ol \psi} 	-\frac{3}{16}\paren{\ola{\p_\mu}}\paren{\ol\psi\gamma^\mu}	&	-\frac{y}{2}\phi \psi^\T	-\frac{3}{16}\paren{\ola{\p_\mu}}\paren{\gamma^\mu \psi}^\T\cr
			\varphi
				&y\ol\psi		&	-y\psi^\T},\\
				\nn
\Gamma_\FB
	&=	\bbordermatrix{
				& h^\perp_{\mu\nu} & \xi_\mu & \sigma & h & \varphi\cr
			\chi^\T&0	&	\frac{1}{4}\paren{{\ol \psi}\gamma^\mu}^\T\sqrt{p^2}	&	\frac{3}{16}\paren{{\ol \psi}\gamma^\mu}^\T\p_\mu	&	-\frac{y}{2}\phi {\ol \psi}^\T+\frac{3}{16}\paren{{\ol \psi}\gamma^\mu}^\T\p_\mu	&	-y{\ol \psi}^\T	\cr
			\ol\chi &0	&	\frac{1}{4}\paren{\gamma^\mu \psi}\sqrt{p^2}	&	\frac{3}{16}\paren{\gamma^\mu\psi}\p_\mu	&	\frac{y}{2}\phi \psi	+\frac{3}{16}\paren{\gamma^\mu\psi}\p_\mu	&	y\psi},\\
			\nn
\Gamma_\FF
	&=	\bmat{
			\Gamma^\text{phys}_\FF&0\\
			0&\Gamma^\text{ghost}_\FF
			},
}
where ${}^\T$ denotes the transposition of the spinor indices and
\al{
\Gamma^\text{phys}_\FF
	&=	\bbordermatrix{
				&\chi&\smash{\ol\chi^\T}\cr
			\chi^\T&0& -\paren{\ola{\Slash D}^\T+y\phi}\cr
			\ol\chi
				& {\Slash D}+y\phi&0
			},\\
\Gamma^\text{ghost}_\FF
	&=	\bbordermatrix{
				&\smash{C^\perp_\nu}&\smash{\bar C^\perp_\nu}&C&\bar C\cr
			C^\perp_\mu&0&	-g^{\mu\nu}\paren{p^2-{R\over 4}}&0&0\cr
			\bar C^\perp_\mu& g^{\mu\nu}\paren{p^2-{R\over 4}} &0&0&0\cr
			C&0&0&0&-\sqbr{\paren{2-{1+\beta\over2}}p^2-{R\over2}}\cr
			\bar C&0&0&\paren{2-{1+\beta\over2}}p^2-{R\over2}&0
			}.
}

\subsection{Explicit form of cutoff functions}\label{cutoff functions}
We write down the cutoff function for the bosons in the non-minimal Higgs-Yukawa model:
\al{
\mc R_\BB
	&=	\bmat{
			\mc R_{h^\perp_{\mu\nu} h^\perp_{\rho\sigma}}&0&0\\
			0	&	\mc R_{\xi_\mu\xi_\nu}&0\\
			0	&	0	&	\mc R_\text{SS}
			},
}
with
\al{
\mc R_{h^\perp_{\mu\nu}h^\perp_{\rho\sigma}}
	&=	\paren{g^{\mu\rho}g^{\nu\sigma}}_\text{sym}{F\over2}R_\Lambda\fn{p^2},\\
\mc R_{\xi_\mu\xi_\nu}
	&=	g^{\mu\nu}{F\over\alpha}R_\Lambda\fn{p^2},\\
		\nn
\mc R_\text{SS}
	&=	\bbordermatrix{
				&\sigma&h&\varphi\cr
			\sigma
				&	\frac{3F}{16}\frac{3-\alpha}{\alpha}R_\Lambda\fn{p^2}
 					&\frac{3F}{16}\frac{\beta-\alpha}{\alpha}K_\Lambda\fn{p^2} 	
					&-\frac{3F_\phi}{4}K_\Lambda\fn{p^2}\cr
			h	&
\frac{3F}{16}\frac{\beta -\alpha}{\alpha}K_\Lambda\fn{p^2}
 	 &	-\frac{F}{16}\frac{3\alpha -\beta ^2}{\alpha}R_\Lambda\fn{p^2}
	 	&  -\frac{3F_\phi}{4} R_\Lambda\fn{p^2}  \cr
			\varphi&
			-\frac{3F_\phi}{4}K_\Lambda\fn{p^2}
 	& -\frac{3F_\phi}{4} R_\Lambda\fn{p^2}
		 &	R_\Lambda\fn{p^2}
	},
}
where we employ the optimized cutoff function~\cite{Litim:2001up}:
\al{
R_\Lambda\fn{p^2}
	&:=	\paren{\Lambda^2-p^2}\theta\paren{\Lambda^2-p^2},\\
K_\Lambda\fn{p^2}
	&:=	\sqrt{p^2+R_\Lambda\fn{p^2}-{R\over 3}}\,\sqrt{p^2+R_\Lambda\fn{p^2}}-\sqrt{p^2-{R\over 3}}\sqrt{p^2}.
}
Note that
\al{
{\p R_\Lambda\fn{p^2}\over\p\Lambda}
	&=	2\Lambda\,\theta\fn{\Lambda^2-p^2},\\
{\p K_\Lambda\fn{p^2}\over\p\Lambda}
	&=	{\Lambda\paren{2\Lambda^2-{R\over 3}}\theta\fn{\Lambda^2-p^2}\over
			\sqrt{p^2+\paren{\Lambda^2-p^2}\theta\fn{\Lambda^2-p^2}}\sqrt{p^2-{R\over 3}+\paren{\Lambda^2-p^2}\theta\fn{\Lambda^2-p^2}}}.
}
More explicitly,
\al{
R_\Lambda\fn{p^2}
	&=	\begin{cases}
		\Lambda^2-p^2,
			&	\\
		0,	&	
		\end{cases}
		&
{\p R_\Lambda\fn{p^2}\over\p\Lambda}
	&=	\begin{cases}
		2\Lambda
			&	\quad\text{for $p^2<\Lambda^2$,}\\
		0	&	\quad\text{for $p^2\geq \Lambda^2$,}
		\end{cases}\\
\nn
K_\Lambda\fn{p^2}
	&=	\begin{cases}
		\sqrt{\Lambda^2-{R\over 3}}\,\sqrt{\Lambda^2}-\sqrt{p^2-{R\over 3}}\sqrt{p^2},
			&	\\
		0,	&	
		\end{cases}
		&
{\p K_\Lambda\fn{p^2}\over\p\Lambda}
	&=	\begin{cases}
		{2\Lambda^2-{R\over 3}\over\sqrt{\Lambda^2-{R\over 3}}}
			&	\quad\text{for $p^2<\Lambda^2$,}\\
		0	&	\quad\text{for $p^2\geq \Lambda^2$.}
		\end{cases}
}
The cutoff function for fermions is given by
\al{
\mc R_\FF
	&=	\bmat{
			\mc R_\text{physical}&0\\
			0&\mc R_\text{ghost}
			},
}
where
\al{
\mc R_\text{physical}
	&=	\bbordermatrix{
				&\chi&\smash{\ol\chi^\T}\cr
			\chi^\T&0&-\ola{\Slash D}^\T\paren{\sqrt{1+{R_\Lambda\fn{p^2+{R\over 4}}\over p^2+{R\over 4}}}-1}\cr
			\ol\chi&\paren{\sqrt{p^2+{R\over 4}+R_\Lambda\fn{p^2+{R\over 4}}\over p^2+{R\over 4}}-1}\Slash D&0\cr
			},\\
			\nn
\mc R_\text{ghost}
	&=
		\bbordermatrix{
				&\smash{C^\perp_\nu}&\smash{\bar C^\perp_\nu}&C&\bar C\cr
			C^\perp_\mu&0&-g^{\mu\nu}R_\Lambda\fn{p^2}&0&0\cr
			\bar C^\perp_\mu&g^{\mu\nu}R_\Lambda\fn{p^2}&0&0&0\cr
			C&0&0&0&-\paren{2-{1+\beta\over2}}R_\Lambda\fn{p^2}\cr
			\bar C&0&0& \paren{2-{1+\beta\over2}}R_\Lambda\fn{p^2} &0
			}.
}
For $\mc R_{\rm physical}$, we have employed the so-called type II cutoff function~\cite{Dona:2012am} in order to give the correct sign of the fermionic quantum corrections for the non-minimal potential $F(\phi^2)$.

\begin{figure}
\begin{center}
\fbox{\includegraphics[width=0.9\textwidth]{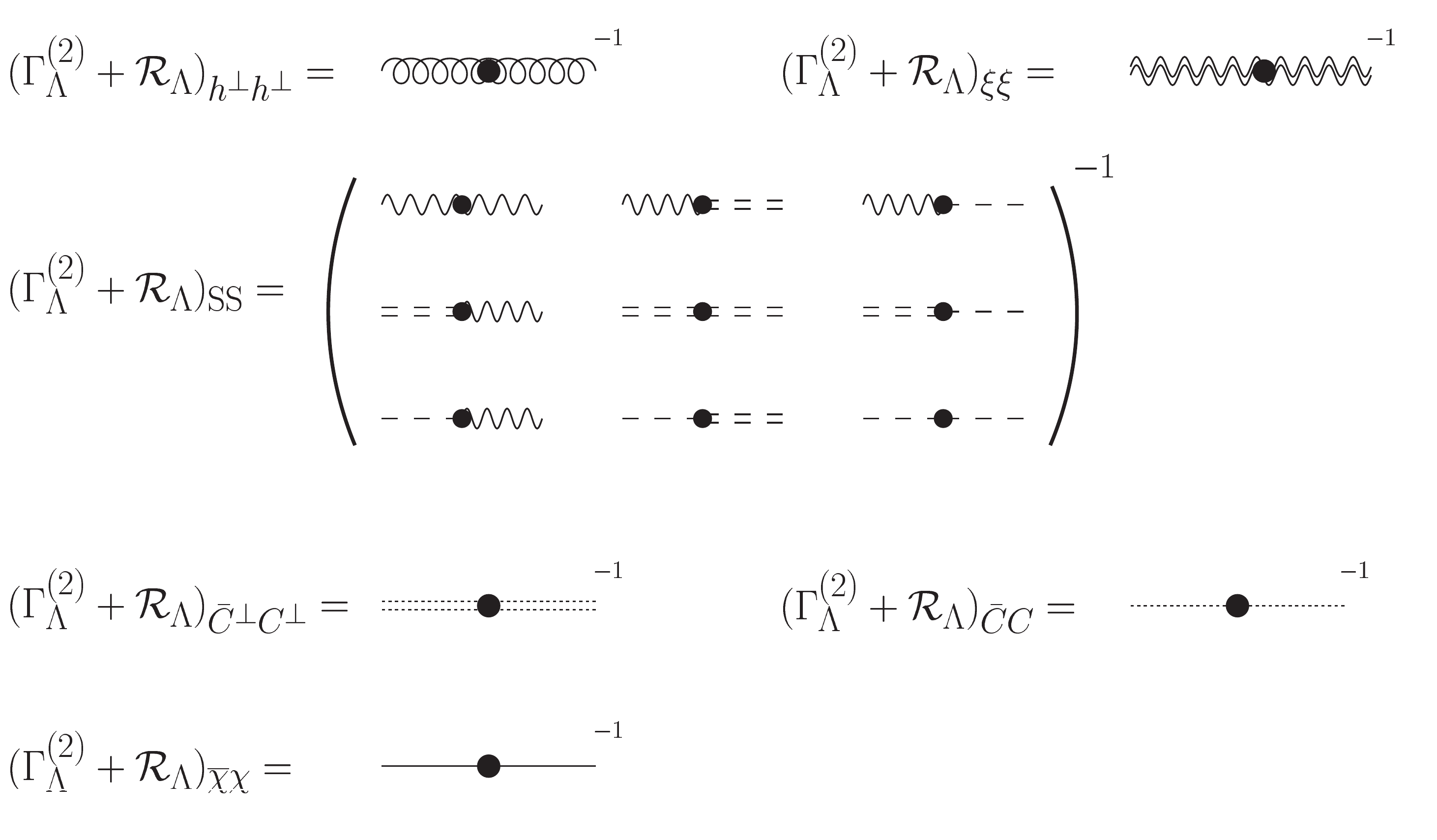}}
\end{center}

\caption{The propagators in the truncated effective action}
\label{propagators}
\end{figure}
\begin{figure}
  \centerline{\hbox{
\includegraphics[width=\textwidth]{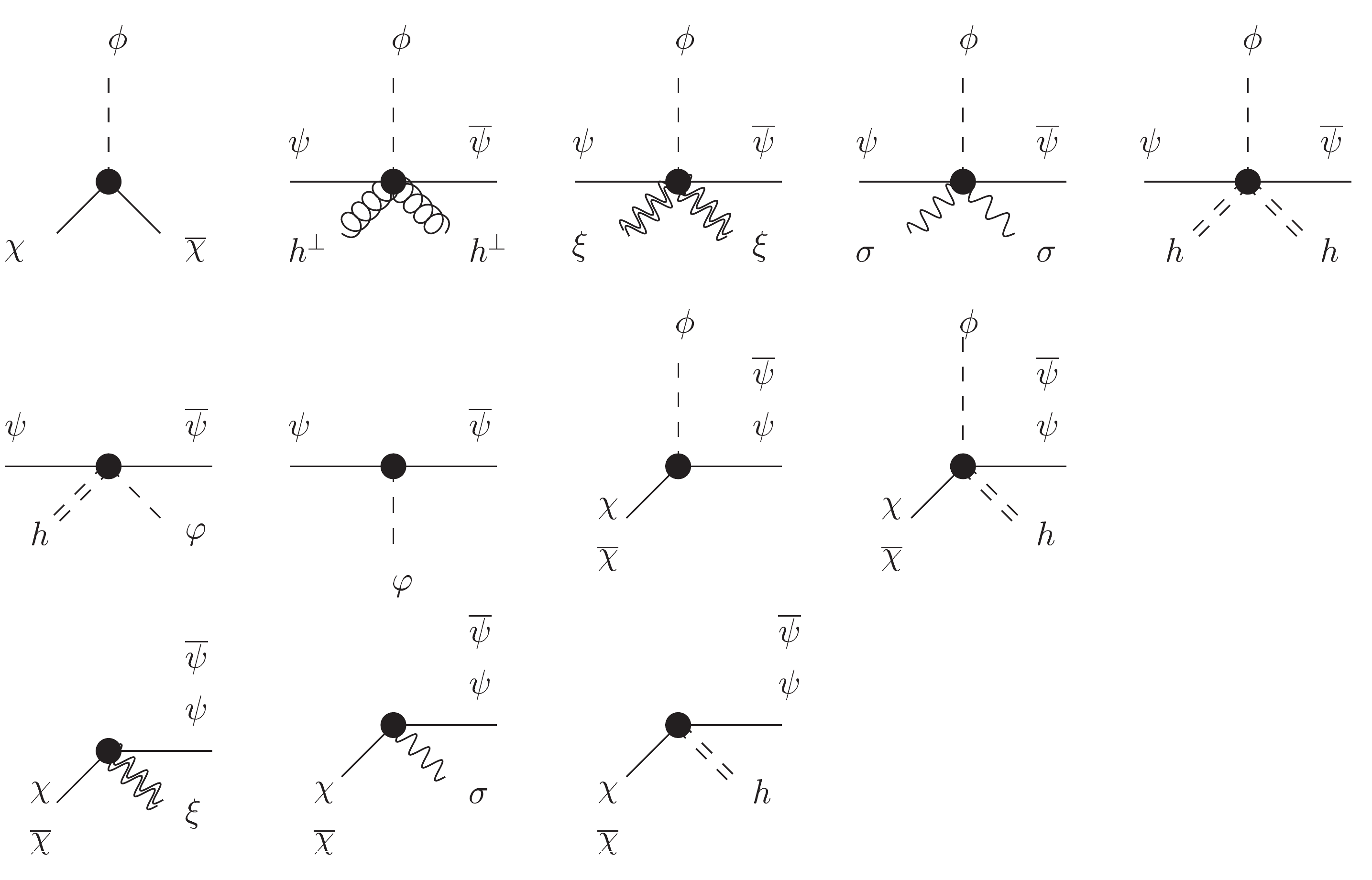}
    }}
\caption{The vertices with fermion in external or internal line}
\label{vertices}
\end{figure}

We have spelled out the two-point and cutoff functions.
From them, we can construct the inverse propagators as in Fig.~\ref{propagators}.
The vertex structures included in $\Gamma_\FB$ and $\Gamma_\BF$ are shown in Fig.~\ref{vertices}.

\section{RG equations}\label{RG equations of system}
\subsection{Computational methods}\label{beta function section}

\begin{figure}
\begin{center}
\includegraphics[width=\textwidth]{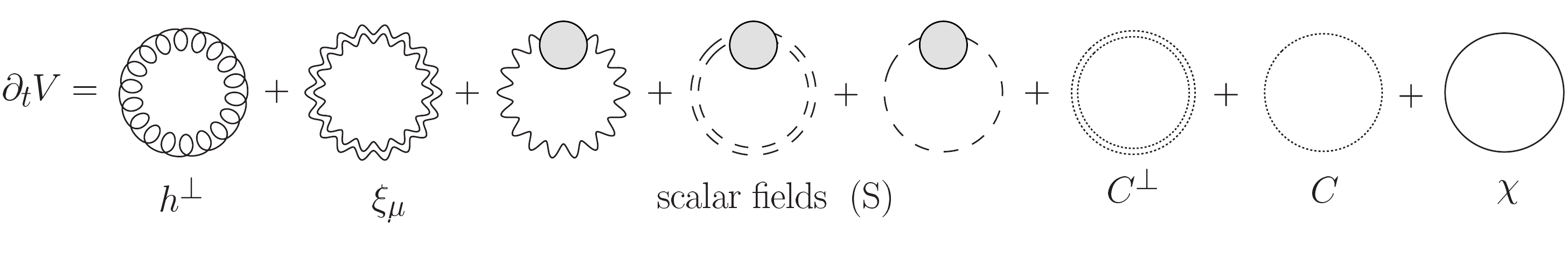}\\
\includegraphics[width=\textwidth]{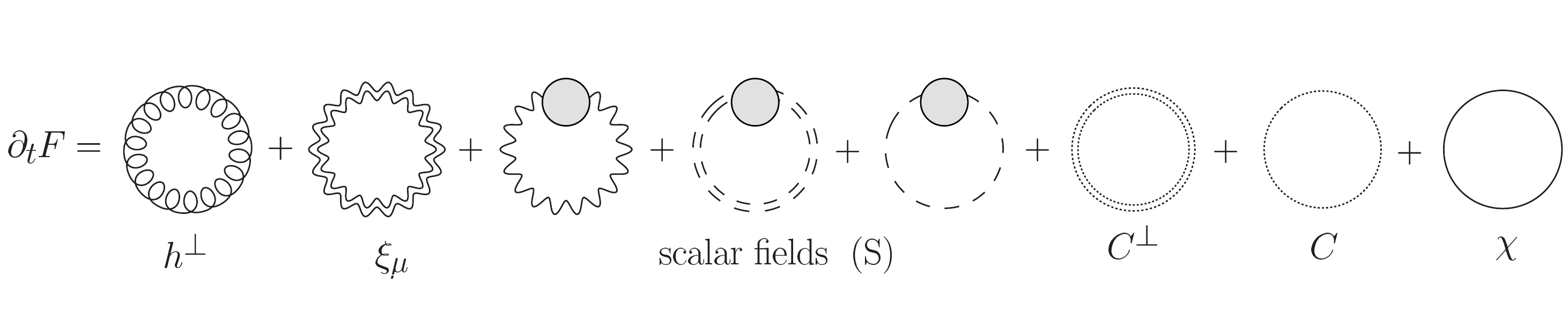}
\end{center}
\caption{One loop contribution to $V$ and $F$.
The gray circle denotes the mixing of scalar fields. }
\label{one loop diagrams}
\end{figure}

There are two methods to compute the beta functions for $V_\Lambda$, $F_\Lambda$ and $Y_\Lambda$.
One is a direct evaluation of the Wetterich equation by algebraic matrix manipulation from the expression
\al{
{\p\over\p\Lambda}\Gamma_\Lambda
	&=	{1\over2}\Tr\bigg[{\mc M}_\BB^{-1}{\p \mc R_\BB\over\p\Lambda}\bigg]\nn
	&\quad
		-{1\over2}\Tr\bigg[\Big({\mc M}_\FF-{\mc M}_\FB {\mc M}_\BB^{-1}{\mc M}_\BF\Big)^{-1}
					\paren{{\p \mc R_\FF\over\p\Lambda}+{\mc M}_\FB {\mc M}_\BB^{-1}\,
						{\p \mc R_\BB\over\p\Lambda}\,
						{\mc M}_\BB^{-1}{\mc M}_\BF}\bigg],
						\label{FRG_total_in_text}
}
where $\mc M_{\Upsilon\Upsilon}=\Gamma_{\Upsilon\Upsilon}+\mc R_{\Upsilon\Upsilon}$ as in Eq.~\eqref{curly M};
explicit forms of the cutoff functions $\mc R_{\Upsilon\Upsilon}$ are given in Sec.~\ref{cutoff functions};
and
$\Gamma_{\Upsilon\Upsilon}$ can be read off as the coefficients of the quadratic terms of the fluctuations $\Upsilon$ in Eq.~\eqref{2daction}.
For the detailed derivation of Eq.~\eqref{FRG_total_in_text}, see Appendix~\ref{frgtreatment}.
We evaluate this expression employing the de-Donder gauge $\alpha=0$, $\beta=1$ after taking the inverse and the trace.

The other is a one-loop diagrammatic computation, obtained by rewriting the Wetterich equation as
\al{
\p_t\Gamma_\Lambda	
	&=	\frac{1}{2}{\rm Tr}\left. \frac{\p_t{\mathcal R}_\Lambda}{\Gamma _\Lambda^{(1,1)}	
					+{\mathcal R}_\Lambda}\right|_{h^\perp h^\perp}
							+\frac{1}{2}{\rm Tr}'\left. \frac{\p_t{\mathcal R}_\Lambda}{\Gamma_\Lambda^{(1,1)}
									+{\mathcal R}_\Lambda}\right|_{\xi \xi}
						+\frac{1}{2}{\rm Tr}''\left. \frac{\p_t{\mathcal R}_\Lambda}{\Gamma _\Lambda^{(1,1)}
									+{\mathcal R}_\Lambda}\right|_{\rm SS}\nn
							&\quad -\left. \Tr \frac{\p_t {\mathcal R} _\Lambda}{\Gamma_\Lambda^{(1,1)}+{\mathcal R}_{\Lambda}}\right|_{\ol\chi \chi}
							-\left. \Tr \frac{\p_t {\mathcal R} _\Lambda}{\Gamma_\Lambda^{(1,1)}+{\mathcal R}_{\Lambda}}\right|_{{\bar C}^\perp C}
							-\left. \Tr \frac{\p_t {\mathcal R} _\Lambda}{\Gamma_\Lambda^{(1,1)}+{\mathcal R}_{\Lambda}}\right|_{\bar C C},
				\label{betafunctions}
}
where we write
\al{
\p_t
	&:=	-\Lambda{\p\over\p\Lambda}
}
and each prime symbol $'$ on the trace denotes a subtraction of a negative eigenvalue of the differential operator from the trace.\footnote{
Since the negative eigenvalues arise from order $R^2$, we ignore it hereafter; see appendix of Ref.~\cite{Codello:2008vh} for details.
}
We see that each term in Eq.~\eqref{betafunctions} can be represented by the corresponding diagram in Fig.~\ref{one loop diagrams}.
Detailed computations are shown in Appendix~\ref{explicitcal}. 
In Appendix~\ref{hkeapp}, we summarizes the values of the heat kernel coefficients used in Appendix~\ref{explicitcal}.

\subsection{Running of $V$ and $F$}
As explained above, we compute the beta function for $V$ and $F$. Its diagrammatic and algebraic derivations are shown in Appendices~\ref{explicitcal} and \ref{frgtreatment}, respectively.
The final results for the beta functions are
\al{
\p_t V
	&=	\frac{\Lambda^4}{192\pi^2}\bigg[ 
			-6-\frac{30V}{\Psi}
			-\frac{6(\Lambda^2\Psi+24\phi^2\Lambda^2F'\Psi'+F\Lambda^2\Sigma_1)}{\Delta} \nn
	&\phantom{=\frac{\Lambda^4}{192\pi^2}\bigg[ }
			+\p_tF\left( \frac{4}{F}+\frac{5\Lambda^2}{\Psi}+ \frac{\Lambda^2\Sigma_1}{\Delta}\right) 
			+\p_tF'\frac{24\phi^2\Lambda^2\Psi'}{\Delta}\bigg]
			+\frac{N_\text{f}}{8\pi^2}\frac{\Lambda^6}{\Sigma_3}, \label{betapotential}\\
\p _tF		
	&=	-\frac{\Lambda^2}{2304\pi^2}	\bigg\{
			150
			+\frac{120\Lambda^2F(3\Lambda^2F-V)}{\Psi^2}  \nn
	&\phantom{=	\frac{\Lambda^2}{2304\pi^2}	\bigg\{}
		-\frac{24}{\Delta}(\Lambda^2\Psi+24\phi^2\Lambda^2F'\Psi'+F\Lambda^2\Sigma_1)\nn
	&\phantom{=	\frac{\Lambda^2}{2304\pi^2}	\bigg\{}
		-\frac{36}{\Delta ^2}\bigg[
			-4\phi^2(6\Lambda^4F'^2+\Psi'^2)\Delta\nn
	&\phantom{=	\frac{\Lambda^2}{2304\pi^2}	\bigg\{}
		\phantom{-\frac{36}{\Delta ^2}\bigg[}
		+4\phi^2\Psi\Psi'\paren{7\Lambda^2F'-V'}\paren{\Sigma_1-\Lambda^2}\nn
	&\phantom{=	\frac{\Lambda^2}{2304\pi^2}	\bigg\{}
		\phantom{-\frac{36}{\Delta ^2}\bigg[}
		+4\phi^2\Sigma_1\paren{7\Lambda^2F'-V'}\paren{2\Psi V'-V\Psi '}
		+(2\Lambda^4\Psi^2 +48\Lambda^4F'\phi^2\Psi \Psi'
			-24\Lambda^4F\phi^2\Psi'^2)\Sigma _2\bigg]\nn
	&\phantom{=	\frac{\Lambda^2}{2304\pi^2}	\bigg\{}
			+\frac{\p_tF}{F}\bigg[
			30	-\frac{10\Lambda^2F (7\Psi	+4V )}{\Psi^2}	 \nn
	&\phantom{=	\frac{\Lambda^2}{2304\pi^2}	\bigg\{}
		\phantom{+\frac{\p_tF}{F}\bigg[}
			+\frac{6}{\Delta ^2} \left(\Lambda^2F\Sigma_1\Delta+ 4\phi^2 V'\Psi'\Delta 
		-24\Lambda^4F\phi^2\Psi'^2\Sigma _2-4\phi^2\Lambda^2F\Psi'\Sigma _1\paren{7\Lambda^2F'-V'}\right)  
		\bigg]\nn
	&\phantom{=	\frac{\Lambda^2}{2304\pi^2}	\bigg\{}
		-\p_tF' \frac{24\Lambda^2\phi^2}{\Delta ^2}
		 \left( \paren{\Lambda^2F'+5V'}\Delta -2\paren{7F'\Lambda^2-V'}\Psi \Sigma_1-12\Lambda^2\Psi\Psi'\Sigma_2 \right) 
		 \bigg\}\nn
	&\quad
		 +\frac{N_\text{f}}{48\pi^2}\frac{\Lambda^4}{\Sigma_3}, \label{betanonminimal}
}
where we employ the notations 
\begin{align}
\label{somedefinitions}
\Psi&:=F\Lambda^2-V, \nn
\Sigma_1&:=\Lambda^2+2V'+4\phi^2V'',\nn
\Sigma_2&:=2F'+4\phi^2F'',\nn
\Sigma_3&:=\Lambda^2+y^2\phi^2,\nn
\Delta&:=12\phi^2\Psi'^2+\Psi\Sigma_1.
\end{align}
In the case of $N_\text{f}=0$, these results~\eqref{betapotential} and \eqref{betanonminimal} agree with those in Ref.~\cite{Narain:2009fy}.
On the other hand, we also reproduce the result in Ref.~\cite{Zanusso:2009bs} when we put 
\cblue{the vanishing non-minimal coupling, i.e.\ $F=\ch\xi_0$ in Eq.~\eqref{all order series}}.
We get the RGE for each coupling constant by
expanding $V(\phi^2)$ and $F(\phi^2)$ into polynomials of the squared scalar field~$\phi^2$:
\al{
V\fn{\phi^2}& =\displaystyle\sum_{n=0}^\infty \ch \lambda_{2n}\phi^{2n}, &
F\fn{\phi^2}&=\displaystyle\sum_{n=0}^\infty \ch \xi_{2n}\phi^{2n}.
	\label{all order series}
}
To investigate the fixed point structure, we define the rescaled dimensionless coupling constants: 
\al{
\lambda_{2n}&:=\ch \lambda_{2n} \Lambda^{2n-4},& \xi_{2n}&:=\ch \xi_{2n} \Lambda^{2n-2}.
	\label{dimensionless couplings}
}
The cutoff $\Lambda$ disappears from RGEs for these dimensionless coupling constants, and there remain the so-called canonical scaling terms:
\al{
\p_t \lambda_{2n} &= -\paren{2n-4}\lambda_{2n} + \text{fluctuations},&\p_t \xi_{2n}&= -\paren{2n-2}\xi_{2n} + \text{fluctuations},
}
where ``fluctuations'' indicate the loop contributions, which are one-loop exact.
Note that the coefficient of the canonical scaling term becomes the dimension of the coupling constant in the LPA.

\subsection{Running of scalar and gravitational coupling constants}\label{explicitbetafunction}
\cred{
We have considered the truncation of full system by restricting to the functional form~\eqref{effectiveaction}. Now we truncate the series in Eq.~\eqref{all order series} up to $\hat\lambda_4$ and to $\hat\xi_2$.
}
We can read off the beta functions for $\xi_0$, $\lambda_0$, $\xi_2$, $\lambda_2$, and $\lambda_4$ from Eqs.~\eqref{betapotential} and \eqref{betanonminimal}.
We show the results in the symmetric phase $\phi=0$:
\al{
\p_t\xi_0&=2\xi_0	-\frac{1}{384\pi^2}\bigg[ 25	
	-\frac{4}{1+2\lambda_2}	-\frac{24\xi_2}{\paren{1 +2\lambda_2}^2}
	+\frac{8\xi_0\paren{7\xi_0-2\lambda_0}}{\paren{\xi_0-\lambda_0}^2}
	\bigg] \nn
	&\quad +\frac{1}{1152\pi^2}\frac{\p_t\xi_0-2{  \xi_0}}{\xi_0} \frac{17\xi_0^2 
		+18\lambda_0 \xi_0 -15\lambda_0^2}{\paren{\xi_0-\lambda_0}^2}
		+\frac{N_\text{f}}{48\pi^2},
		\label{xi0 RGE}
 \\
\p_t \lambda_0&=4\lambda_0-\frac{1}{32\pi^2}\bigg[ 2	+\frac{1}{1+2\lambda_2}	
								+\frac{6\lambda_0}{\xi_0 - \lambda_0} \bigg]
		+\frac{\p _t \xi_0-2{  \xi_0}}{96\pi^2 \xi_0}\frac{5\xi_0-2\lambda_0}{\xi_0 -\lambda_0}
		+\frac{N_\text{f}}{8\pi^2},
		\label{lambda0 RGE}
}
\al{
\p_t\xi_2
	&=	-\frac{1}{576\pi^2}\bigg[
			\frac{1+2\lambda_2}{\xi_0-\lambda_0}\paren{
				9+\frac{39\xi_0}{\xi_0-\lambda_0}
				+\frac{60\xi_0^2}{\paren{\xi_{\cred 0}-\lambda_0}^2}
				}
			+\frac{3\paren{3+32\xi_2}}{\xi_0-\lambda_0}	
			-\frac{6\xi_0\paren{11+2\xi_2}}{\paren{\xi_0-\lambda_0}^2}
			 \nn
	&\phantom{=	-\frac{1}{576\pi^2}\bigg[} 
		-\frac{60\xi_0^2\paren{1+2\xi_2}}{\paren{\xi_0-\lambda_0}^3}	
		+\frac{216\xi_2\paren{1+2\xi_2}^2}{\paren{1+2\lambda_2}^3\paren{\xi_0-\lambda_0}}
		+\frac{
			9\sqbr{\lambda_0\paren{5-2\xi_2}-2\xi_0\paren{1+2\xi_2}}
			\paren{1+2\xi_2}
			}{\paren{1+2\lambda_2}\paren{\xi_{\cred 0}-\lambda_0}^2}\nn
	&\phantom{=	-\frac{1}{576\pi^2}\bigg[}  
		+\frac{27\paren{1+2\xi_2}\paren{1-10\xi_2-16\xi_2^2}}{\paren{1+2\lambda_2}^2\paren{\xi_0-\lambda_0}}
		+\frac{108\xi_0 \xi_2\paren{1+2\xi_2}^2}{\paren{1+2\lambda_2}^2\paren{\xi_0-\lambda_0}^2}
		+\frac{72{   \lambda}_4}{\paren{1+2\lambda_2}^2} \frac{1+12\xi_2+2\lambda_2}{1+2\lambda_2} \bigg]\nn
		&\quad+\frac{\p_t \xi_0-2\xi_0}{1152\pi^2 \xi_0}\bigg[
				\frac{1+2\lambda_2}{\xi_0 -\lambda_0}\bigg(
					3
					+\frac{18\xi_0}{\xi_0 -\lambda_0}	
					+\frac{20\xi_0^2}{\paren{\xi_0-\lambda_0}^2}
					\bigg)
			+\frac{15\xi_2}{\xi_0}	
			-\frac{6\paren{1+\xi_2}}{\xi_0-\lambda_0}
			-\frac{10\xi_0\paren{3+4\xi_2}}{\paren{\xi_0-\lambda_0}^2}\nn
		&\phantom{\quad+\frac{\p_t \xi_0-2\xi_0}{1152\pi^2 \xi_0}\bigg[}
			-\frac{20\xi_0^2\paren{1+2\xi_2}}{\paren{\xi_0-\lambda_0}^3}
		 	-\frac{3\sqbr{\lambda_0-\xi_0\paren{5-4\xi_2}}\paren{1+2\xi_2}}
				{\paren{1+2\lambda_2}\paren{\xi_0-\lambda_0}^2} 
			+\frac{36\xi_0 \xi_2\paren{1+2\xi_2}^2}
				{\paren{1+2\lambda_2}^2\paren{\xi_0-\lambda_0}^2}  
			\bigg] \nn
		&\quad +\frac{\p_t\xi_2}{1152\pi^2\xi_0}\bigg[ -15	
		+\frac{54\xi_0}{\xi_0-\lambda_0}	
			+\frac{20\xi_0^2}{\paren{\xi_0-\lambda_0}^2}	
			-\frac{6\xi_0\paren{7+2\xi_2}}
				{\paren{1+2\lambda_2}\paren{\xi_0-\lambda_0}} 
			-\frac{144\xi_0 \xi_2\paren{1+2\xi_2}}
				{\paren{1+2\lambda_2}\paren{\xi_0-\lambda_0}}  \bigg]\nn
		&\quad
			-\frac{N_\text{f} y^2}{48\pi^2},
			\label{xi2 RGE}\\
\p_t \lambda_2 &= 2 \lambda_2 - \frac{1}{48 \pi ^2} 
	\bigg[ 
		 \frac{9  \lambda_0 \paren{1+ 2 \xi_2} }{2\paren{\xi_0- \lambda_0}^2} 
			- \frac{9\paren{2\lambda_0- \xi_0}\paren{1+ 2  \xi_2}^2 }{2\paren{1+ 2  \lambda_2}\paren{\xi_0-\lambda_0}^2}			
			- \frac{9 \paren{1+2\xi_2}^2 }{2 \paren{1+2\lambda_2}^2  \paren{\xi_0-\lambda_0}} - \frac{18 \lambda_4 }
			{\paren{1+2\lambda_2}^2}
		\bigg]
\nn
		&\quad+ \frac{\p_t \xi_0-2\xi_0}{96 \pi ^2\xi_0} 
	\bigg[
		- \frac{2  \xi_2}{\xi_0} + \frac{3  \xi_0 \paren{1+2\xi_2} }{2 \paren{\xi_0-\lambda_0}^2}
		- \frac{3 \xi_0 \paren{1+2\xi_2}^2 }{2 \paren{1+2\lambda_2} \paren{\xi_0-\lambda_0}^2}	
	 \bigg] \nn
	&\quad+  \frac{1}{96 \pi ^2} \frac{\p_t  \xi_2}{ \xi_0} 
	\bigg[ 
		2 -\frac{3  \xi_0}{\xi_0- \lambda_0}
		+\frac{6  \xi_0 \paren{1+2\xi_2} }{\paren{1+2\lambda_2} \paren{\xi_0-\lambda_0}} 
	\bigg] 
	-\frac{N_\text{f} y^2}{8\pi^2},
		\label{lambda2 RGE}
}
\al{
\p _t  \lambda_4
	&=	- \frac{1}{48 \pi ^2}\Bigg[
			\frac{9}{4 \paren{\xi_0- \lambda_0}^2 }\Bigg(
				5\paren{1+2\lambda_2}\paren{1+4 \xi_2}
				-\paren{1+2\xi_2}\paren{21+62 \xi_2}  
				+ \frac{33 \paren{1+2\xi_2}^3}{1+2\lambda_2}\nn
	&\phantom{=	- \frac{1}{48 \pi ^2}\Bigg[}
		\phantom{\frac{9}{4 \paren{\xi_0- \lambda_0}^2 }\Bigg(}
			- \frac{\paren{1+2\xi_2}^3 \paren{23+24 \xi_2}}{\paren{1+2\lambda_2}^2}
			+ \frac{6 \paren{1+2\xi_2}^4}{\paren{1+2\lambda_2}^3}\Bigg)\nn
	&\phantom{=	- \frac{1}{48 \pi ^2}\Bigg[}
		+ \frac{9 \xi_0\paren{\xi_2- \lambda_2}^2}{\paren{\xi_0-\lambda_0}^3} 
			\paren{6 \frac{\paren{1+2\xi_2}^2}{\paren{1+2\lambda_2}^2} 
				- 10\frac{1 + 2  \xi_2}{1 + 2  \lambda_2}+ 5}
		-\frac{72 \lambda_2 \lambda_4\paren{1+2\xi_2}\paren{1 - 4 \lambda_2 + 6 \xi_2 }}
		{\paren{\xi_0-\lambda_0}\paren{1+2\lambda_2}^3} \nn
	&\phantom{=	- \frac{1}{48 \pi ^2}\Bigg[} 
		+\frac{9 \xi_0 \lambda_4}{\paren{\xi_0-\lambda_0}^2} 		
			\paren{6 \frac{\paren{1+2\xi_2}^2}{\paren{1+2\lambda_2}^2} 
				- 8 \frac{1 + 2   \xi_2}{1 + 2   \lambda_2}+ 3}
		+\frac{216  \lambda_4^2}{\paren{1+2\lambda_2}^3}  
		\Bigg] \nn
	&\phantom{=	- \frac{1}{48 \pi ^2}\Bigg[}
		+ \frac{\p_t \xi_0 -2\xi_0}{96\pi^2\xi_0}\Bigg[
			\frac{2 \xi_2^2}{\xi_0^2} 
			+\frac{3\xi_0\paren{\xi_2-\lambda_2}^2}
					{\paren{\xi_0-\lambda_0}^3} 
				\paren{
					6\frac{\paren{1+2\xi_2}^2}{\paren{1+2\lambda_2}^2} 
					- 10 \frac{1 + 2  \xi_2}{1 + 2  \lambda_2}
					+ 5
					} \nn 
	&\phantom{=	- \frac{1}{48 \pi ^2}\Bigg[}
		\phantom{+ \frac{\p_t \xi_0 -2\xi_0}{96\pi^2\xi_0}\Bigg[}
			+ \frac{3 \xi_0  \lambda_4}{\paren{\xi_0-\lambda_0}^2} 
				\paren{6\frac{\paren{1+2\xi_2}^2}{\paren{1+2\lambda_2}^2} 
				- 8 \frac{1 + 2  \xi_2}{1 + 2   \lambda_2}+ 3}
			\Bigg]\nn
	&\quad+\frac{1}{96\pi^2}\frac{\p_t \xi_2}{\xi_0}\Bigg[
			-\frac{2\xi_2}{ \xi_0}
	 		-\frac{24\xi_0\lambda_4\paren{1-4 \lambda_2+6\xi_2}}{\paren{1+2\lambda_2}^2\paren{\xi_0-\lambda_0}}
	\nn 
	&\phantom{\quad+\frac{1}{96\pi^2}\frac{\p_t \xi_2}{\xi_0}\Bigg[}
			- \frac{3  \xi_0\paren{\xi_2-\lambda_2}}
					{\paren{\xi_0-\lambda_0}^2} 
				\paren{12\frac{\paren{1+2\xi_2}^2}{\paren{1+2\lambda_2}^2} 
					- 21 \frac{1 + 2  \xi_2}{1 + 2  \lambda_2}+ 10}
				\Bigg]
	+\frac{N_\text{f} y^4}{8\pi^2}.
	\label{lambda4 RGE}
}
The last term of each beta function is coming from the fermionic fluctuation.
The others agree with the results in Ref.~\cite{Narain:2009fy}.
When $y=0$, we see that the loop of $\psi$ contributes only to the beta functions of $\xi_0$ and $\lambda_0$.

In Ref.~\cite{Zanusso:2009bs}, the authors have studied the Higgs-Yukawa model without the non-minimal coupling $\xi_2=0$. We have checked that when $\xi_2=0$, our RG equation for the scalar potential~\eqref{betapotential} reduces to theirs, namely the first line of Eq.~(4) in Ref.~\cite{Zanusso:2009bs}, if we impose that the dimension\emph{ful} gravitational coupling constant $\hat\xi_0$ does not run, $\p_t\hat\xi_0=\p_t\xi_0-2\xi_0=0$, in the right hand side of the RG equation. (We write $\xi_0=1/16\pi\tilde G$ where $\tilde G$ is the dimensionless Newton constant.)
Similarly, we can see that the RG equations for $\lambda_0$~\eqref{lambda0 RGE}, for $\lambda_2$~\eqref{lambda2 RGE}, and for $\lambda_4$~\eqref{lambda4 RGE} reduce to Eq.~(6) in Ref.~\cite{Zanusso:2009bs} if we put $\lambda_0=\xi_2=0$ and $\p_t\hat\xi_0=0$.

\subsection{Running of Yukawa coupling}
\begin{figure}
  \centerline{\hbox{
\includegraphics[width=\textwidth,bb= 0 0 1006 569]{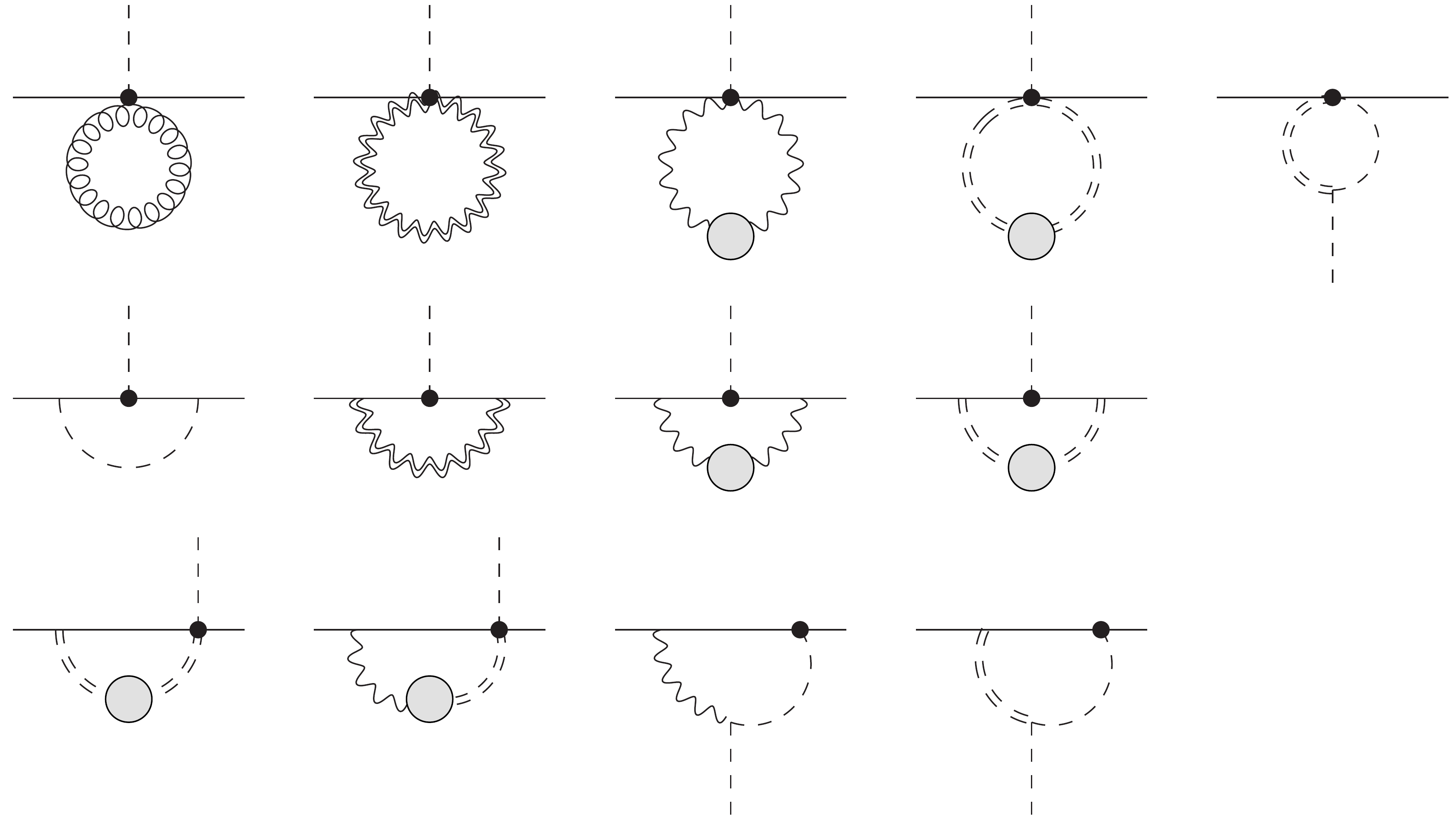}
\put(-440,240){(I)}
\put(-350,240){(II)}
\put(-260,240){(III)}
\put(-170,240){(IV)}
\put(-80,240){(V)}
\put(-440,150){(VI)}
\put(-350,150){(VII)}
\put(-260,150){(VIII)}
\put(-170,150){(IX)}
\put(-440,80){(X)}
\put(-350,80){(XI)}
\put(-260,80){(XII)}
\put(-170,80){(XIII)}
    }}
\caption{The corrections to the Yukawa coupling constant in our truncation. The black dot and the gray circle denote the Yukawa interaction vertex and the mixing of scalar fields, respectively. }
\label{yukawacorrections}
\end{figure}
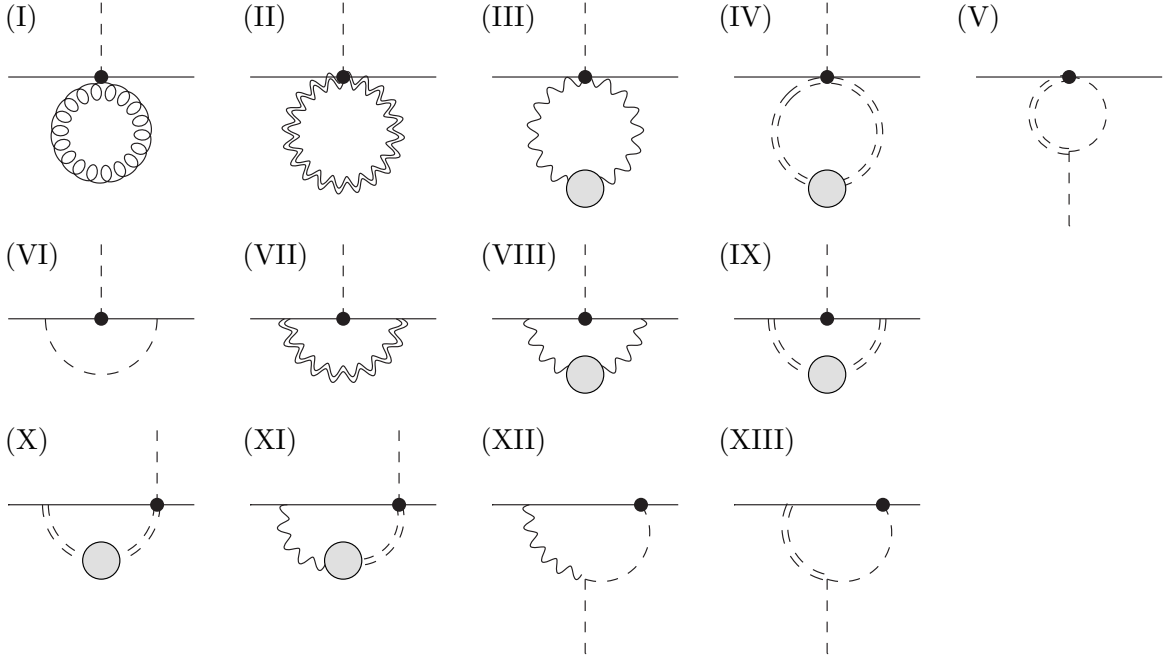

By the same two methods described in Sec.~\ref{beta function section}, we obtain the RGE for $Y$, and read off the beta function for the Yukawa coupling constant $y$ in the symmetric phase $\phi=0$:
\al{
\p_ty
	&=	\frac{5y\Lambda^6}{32\pi^2}\left( \frac{\p_t \ch \xi_0}{6} -\ch \xi_0 \right)I[2,0,0]\nn
	&\quad+\frac{y\Lambda^6}{32\pi^2}\bigg[ 
		24\left( \ch\xi_2 -\frac{\p_t \ch\xi _2}{6}\right) I[1,1,0]
		-\left(\ch\xi_0-\frac{\p_t\ch\xi_0}{6}\right) I[2,0,0]	 - \cred{12}\,C\left( \ch\xi_0-\frac{\p_t \ch\xi_0}{6}\right) I[2,1,0]	 \nn
	&\phantom{\quad+\frac{y\Lambda^6}{32\pi^2}\bigg( }
		-\cred{12}\,CI[1,2,0]
	\bigg]\nn
	&\quad
		-\frac{y^3\Lambda^6}{16\pi^2}\paren{I[0,1,2]+I[0,2,1]}
		-\frac{y\Lambda^8}{128\pi^2}\left[ I[1,0,2] + \paren{\ch\xi_0-\frac{\p_t \ch\xi_0}{8}} I[2,0,1] \right]\nn
	&\quad
		+\frac{3y\Lambda^8}{40\pi^2}\bigg[ I[1,0,2] + \left( \ch\xi_0-\frac{\p_t\ch\xi_0}{7} \right)I[2,0,1] - \frac{1}{2\Lambda^2}I[1,0,1] \bigg]\nn
	&\quad
		-\frac{3y\Lambda^8}{20\pi^2}\bigg[
	-\left( \ch\xi_0 -\frac{\p_t \ch\xi_0}{7} \right) CI[2,1,1]	 +\left( \ch\xi_2 -\frac{\p_t \ch\xi_2}{7} \right)I[1,1,1]  \nn
	&\phantom{\quad-\frac{y\Lambda^8}{16\pi^2}\bigg(}
		-C\left( I[1,2,1] + I[1,1,2]	- \frac{1}{2\Lambda^2} I[1,1,1] \right)
		\bigg], \label{betayukawa}
}
where $C=\ch\xi_2\Lambda^2-\ch\lambda_2$ and 
\al{
I[n_g,n_b,n_f]:=\left. \frac{1}{\Psi^{n_g} \Sigma_1^{n_b}\Sigma_3^{n_f}}\right|_{\phi=0}
			=\frac{1}{(\ch\xi_0\Lambda^2-\ch\lambda_0)^{n_g}(\Lambda^2 +2\ch \lambda_2)^{n_b}(\Lambda^2)^{n_f}}.
}
This is one of our main results.

The first term in the beta function corresponds to the diagram (I) in Fig.~\ref{yukawacorrections}.
The second term includes the diagrams (III), (IV) and (V).
The term in proportion to $y^3$ corresponds to the diagram (VI).
The terms in fourth and fifth lines in the beta function correspond to (VIII)(IX) and (X)(XI), respectively.
The last terms correspond to (XII) and (XIII). 

Using the dimensionless rescaled coupling constants $\lambda_{2n}$, $\xi_{2n}$ introduced in Eq.~\eqref{dimensionless couplings}, we can easily rewrite the beta function of $y$ by the replacements
\al{
\p_t\hat\lambda_{2n}
	&=	{1\over\Lambda^{2n-4}}\sqbr{\p_t\lambda_{2n}+\paren{2n-4}\lambda_{2n}},&
\p_t\hat\xi_{2n}
	&=	{1\over\Lambda^{2n-2}}\sqbr{\p_t\xi_{2n}+\paren{2n-2}\xi_{2n}}.
		\label{to dimensionless}
}
Since the Yukawa coupling constant is dimensionless and its canonical scaling term vanishes in its beta function in the LPA, we have been omitting the hat $\hat{\mbox{ }}$ for $y$.

Let us try to put $\lambda_0=\xi_2=0$ as in the end of Sec.~\ref{explicitbetafunction}.
If we impose that the dimensionful constant does not run, $\cred{\p_t\hat\xi_0=\Lambda^2(\,}\p_t\xi_0-2\xi_0\cred{)}=0$, we obtain\footnote{The difference of overall sign in the beta function is due to the sign convention for the dimensionless scale~$t$. Recall that in our notation, $t=-\log\fn{\Lambda/\Lambda_0}$.}
\al{
\dot y
	&:=	-\beta _y
	=	\frac{y^3(1+\lambda_2)}{8\pi^2\paren{1+2\lambda_2}^2}
		+\tilde Gy{29-4\lambda_2\paren{\cblue1-\cblue5\lambda_2}\over20\pi\paren{1+2\lambda_2}^2}.
		\label{Yukawa RGE to compare}
}
This is to compare with Eq.~(6) in Ref.~\cite{Zanusso:2009bs}.
We see that the first term, which corresponds to the diagram (VI) in Fig.~\ref{yukawacorrections}, agrees each other, while the second term does not.
To study the fixed-point structure, we are rather interested in the limit where dimensionless coupling constant does not run $\p_t\cred{\xi_0}\to 0$, which results in 
\al{
\dot y
	&:=	-\beta _y
	=	\frac{y^3\paren{1+\lambda_2}}{8\pi^2\paren{1+2\lambda_2}^2} 
		+\tilde G y \frac{2395 + 4\lambda_2 \paren{\cblue{347} +\cblue{315}\lambda_2}}{1120\pi \paren{1+2\lambda_2}^2}
		+\mathcal O\fn{{\tilde G}^2}.
}
However, the dimensionless cosmological constant $\lambda_0$ is not vanishing at the UV fixed point, and we will rely on the numerical computation in the next section.

\section{Numerical Analysis}\label{numericalan}

\subsection{Fixed Point structure}
The fixed points are defined by vanishing beta functions
$\beta_i\fn{g^*}=0$ at which RG flows completely stop.
To study the behavior of the RG flow near the fixed point $g^*$, let us consider the linearized flow equations.
Let $N$ be the dimension of our (truncated) coupling space.
We expand the beta function around $g^*$,
\al{
\beta_i\fn{g} = \beta_i\fn{g^*} + \sum_{j=1}^N\left. \frac{\p \beta_i}{\p g_j}\right|_{g=g^*}\paren{g_j-g_j^*} +\cdots.
}
Using $\beta_i\fn{g^*}=0$ and neglecting the higher order terms in $v:=g-g^*$, we obtain the linearlized RGE 
\al{\label{linearbetaeq}
\p_t v_i = \sum_{j=1}^N\left. \frac{\p \beta_i}{\p g_j}\right|_{g=g^*} v_j.
}
Let us diagonalize the matrix
\al{
M_{ij}:=\left.{\p \beta_i\over\p g_j}\right|_{g=g^*}
	\label{beta matrix}
}
by a constant matrix $V$ so that
\al{
\sum_{i,j=1}^N\paren{V^{-1}}_{li}\left.{\p\beta_i\over\p g_j}\right|_{g=g^*}V_{jk}
	&=	\theta_k\delta_{lk},
}
where $k$ is not summed. That is, the $k$th eigenvalue of $M$ is $\theta_k$, and the corresponding eigenvector is $V^{(k)}=\paren{V_{jk}}_{j=1,\dots,N}$:
\al{
MV^{(k)}=\theta_kV^{(k)}.
}
Now Eq.~\eqref{linearbetaeq} reduces to
\al{
\p_t \kappa_i= \theta _i \kappa_i,
	\label{kappa equation}
}
where the index $i$ is not summed and we have written $v_i=\sum_{j=1}^NV_{ij}\kappa_j$.
The solutions to Eq.~\eqref{kappa equation} are
\al{
\kappa_i (t) = C_i e^{\theta_i t},
	\label{linearlizedoperator}
}
where $C_i$ are constants. 
When we recover the original dimensionless coupling constants $g_i$, Eq.~\eqref{linearlizedoperator} reads
\al{\label{linearfloworiginal}
g_i(t) = g_i^* +\displaystyle \sum_{j=1}^N  V_{ij}C_j \left( \frac{\Lambda_0}{\Lambda} \right)^{\theta_j},
}
which becomes Eq.~\eqref{fixedrg} with $\zeta_{ij}=C_jV_{ij}$. 
In general, a non-zero $\im\fn{\theta_i}$ implies that the corresponding coupling $g_i$ is mixed with other couplings in the RG flow from UV to IR, $\Lambda\to0$, i.e.\ $t\to\infty$.
Let us see three cases in turn:
\begin{itemize}
\item For the directions with $\re\fn{\theta_i}>0$, we see that $\kappa_i$ grow when we increase $t$ in the flow from UV to IR. Then $g_i$ become the couplings of the relevant operators, and the factor $C_i$ become physical free parameters. When we vary ratios of $C_i$, the direction of the flow to IR changes, and we get a different IR physics.
\item For the directions with $\re(\theta_i)=0$, the solutions~\eqref{linearlizedoperator} generally become oscillatory, and the corresponding operators are marginal.
\item For the directions with $\re(\theta_i)\,{\cred<}\,0$, the solutions shrink to the UV fixed point, and hence they are the coupling of the irrelevant operators.
\end{itemize}
The relevant operators span the hypersurface called the renormalized trajectory or the UV critical surface, and the number of such operators gives its dimension. 


\subsection{Pure gravity}
We first revisit the pure gravity case obtained in Refs.~\cite{Percacci:2003jz,Narain:2009fy,Narain:2009gb}.
The beta functions for the dimensionless gravitational coupling $\xi_0$ and the dimensionless cosmological constant $\lambda_0$ become
\al{
\beta_{\xi_0} &= 2\xi_0 -\frac{1}{384\pi^2}\left[21 + \frac{8\xi_0(7\xi_0-2\lambda_0)}{\paren{\xi_0-\lambda_0}^2} \right]
								+\frac{\p_t \xi_0 -2\xi_0}{1152\pi^2 \xi_0} \frac{17\xi_0^2 +18\xi_0\lambda_0 -15\lambda_0^2}{\paren{\xi_0-\lambda_0}^2},\\
\beta _{\lambda_0}&= 4\lambda_0 - \frac{1}{32\pi^2} \left[3+ \frac{6\lambda_0}{\xi_0-\lambda_0}\right] +\frac{\p_t \xi_0-2\xi_0}{96\pi^2\xi_0}\frac{5\xi_0-2\lambda_0}{\xi_0-\lambda_0}.
}
Solving the coupled equation $\beta_{\xi_0} =0$ and $\beta _{\lambda_0}=0$, we find the non-trivial fixed point:
\al{
\xi_0^*
	&=	2.38 \times 10^{-2},	&
\lambda_0^*
	&=	8.62\times 10^{-3}.
} 
Around the fixed point, the matrix~\eqref{beta matrix} becomes:
\al{
M=
\left.
\pmat{
\displaystyle \frac{\p \beta_{\xi_0}}{\p \xi_0} & \displaystyle\frac{\p \beta_{\xi_0}}{\p \lambda_0}\\[10pt]
\displaystyle  \frac{\p \beta_{\lambda_0}}{\p \xi_0} & \displaystyle \frac{\p \beta_{\lambda_0}}{\p \lambda_0}
}\right| _{\xi_0=\xi_0^*\atop \lambda_0=\lambda_0^*}
=
\pmat{
4.79429 & -6.42602\\
2.33613 & -0.505996
}
.
}
The eigenvalues for this matrix are $\theta_{1,2}=2.14414 \pm 2.82644 i$. 
We see that $\xi_0$ and $\lambda_0$ are the relevant coupling constants around the UV fixed point. The corresponding eigenvectors are
\al{
V^{(1)},V^{(2)}
	&=
\pmat{
0.856378\\
0.353177 \pm 0.376673 i
},	&
V	&=	\pmat{
			0.856378	&	0.856378\\
			0.353177 + 0.376673 i&0.353177- 0.376673 i}.
}
That is,
\al{
\xi_0(t) 
	&=	\xi_0^* + 0.856378\paren{\Lambda_0\over\Lambda}^{2.14414}\sqbr{
			A\cos\fn{\ln\paren{\Lambda_0\over\Lambda}^{2.82644}}
			+B\sin\fn{\ln\paren{\Lambda_0\over\Lambda}^{2.82644}}
			},\\
\lambda_0(t) 
	&= \lambda_0^* + \paren{\Lambda_0\over\Lambda}^{2.14414}\Bigg[
			\paren{0.353177A+0.376673B}\cos\fn{\ln\paren{\Lambda_0\over\Lambda}^{2.82644}}\nn
	&\phantom{= \lambda_0^* + \paren{\Lambda_0\over\Lambda}^{2.14414}\Bigg[}
			+\paren{-0.376673A+0.353177B}\sin\fn{\ln\paren{\Lambda_0\over\Lambda}^{2.82644}}
			\Bigg],
}
where $A:=C_1+C_2$ and $B:=i\paren{C_1-C_2}$ are real constants that are free parameters of the asymptotically safe theory.
We see that the two relevant couplings mix with each other in the RG flow to IR scales.

\subsection{Scalar-gravity model}\label{scalar-gravity model}
Next we turn to the extension of the system with the neutral scalar field~\cite{Percacci:2003jz,Narain:2009fy,Narain:2009gb}.
This section is still a review.
In truncated theory space $g_i=\{ \xi_0, \lambda_0, \xi_2, \lambda_2, \lambda_4 \}$,
we find the fixed point
\al{
\xi_0^*&=2.38 \times 10^{-2},& 
\lambda_0^*&=8.62 \times 10^{-3},& 
\xi_2^*&=0,& 
\lambda_2^*&=0,& 
\lambda_4^*&=0.
	\label{Gaussian-matter fixed point}
}
The gravitational coupling constants have the same non-trivial fixed point, while the matter fixed point is trivial, i.e.\ Gaussian.
The fixed point~\eqref{Gaussian-matter fixed point} is called the Gaussian-matter fixed point.\footnote{\cred{
It could be that the Gaussian-matter fixed point is a special property of the present truncation with LPA: If we take into account the higher-derivative matter self-interactions, which are induced by the gravitational fluctuations, then the matter self interaction might become non-vanishing at the fixed point~\cite{Eichhorn:2012va}. 
}}
The matrix~\eqref{beta matrix} becomes
\al{
M=
\pmat{
4.85544	&	-6.51993		&	0.00766245	&	-0.00262748	&	0 \\
2.40051	&	-0.570309		&	0.00234951	&	0.0055494	&	0 \\
0		&	0			&	2.85544		&	\cblue{-6.51993}		&	-0.0157649 \\
0		&	0			&	2.40051		&	-2.57031		&	0.0332964 \\
0		&	0			&	0			&	0			&	-2.62692
},
}
and the eigenvalues and the corresponding eigenvectors are
\al{
\theta_{1,2}&= 2.143 \pm 2.879 i,&
V^{(1)},V^{(2)}&=
\pmat{
0.8549\\
0.3557 \pm 0.3776i \\
0\\
0\\
0
},
 \\
\theta_{3,4}&= 0.143 \pm 2.879i, &
V^{(3)},V^{(4)}&=
\pmat{
\paren{-1.8059 \pm 0.731i}\times 10^{-3}\\
\paren{3.0723 \pm 1.0763i}\times 10^{-3}\\
0.3557 \pm 0.3776i\\
0.8549\\
0
},
\\
\theta_{5}&= -2.627, &
V^{(5)}&=
\pmat{
2.0687 \times 10^{-5}\\
2.7445 \times 10^{-5}\\
-1.3805 \times 10^{-2}\\
-1.3542 \times 10^{-2}\\
0.999813
}.
}
Several comments are in order:
\begin{itemize}
\item Since the vectors $V^{(1)}$ and $V^{(2)}$ have the values at only first and second rows, these vectors correspond to the mixing between $\xi_0$ and $\lambda_0$. 
These coupling constants are relevant as their critical exponents are positive: $\re\theta_1>0$ and $\re\theta_2>0$.
We see that the impact of scalar fluctuations to the gravitational couplings is not large since the values of the critical exponents $\theta_{1,2}$ hardly change by its inclusion.
\item Although the vectors $V^{(3)}$ and $V^{(4)}$ include the mixing of $\xi_0$, $\lambda_0$, $\xi_2$ and $\lambda_2$, the contributions from the gravitational couplings $\xi_0$ and $\lambda_0$ are smaller than those from $\xi_2$ and $\lambda_2$.
Therefore they are mainly $\xi_2$ and $\lambda_2$.
These coupling constants are relevant as their critical exponent is positive: $\re\theta_3>0$ and $\re\theta_4>0$.
Note that the non-minimal coupling constant $\xi_2$ is marginal at the trivial fixed point $g^*=0$, and hence, the gravitational effects have made it relevant.
\item The scalar quartic coupling $\lambda_4$ is irrelevant as its critical exponent is negative: $\re\theta_5<0$.
Although $\lambda_4$ is marginal at the trivial fixed point, the gravitational effects make it irrelevant at the UV fixed point.
\end{itemize}
In this truncated theory space, the UV critical surface is spanned by the operators with $\xi_0$, $\lambda_0$, $\xi_2$ and $\lambda_2$. 
Hence, these coupling constants are physical free parameters~\cite{Percacci:2003jz,Narain:2009fy}.
In next part we will see that the fermionic fluctuation makes $\xi_2$ and $\lambda_2$ irrelevant so that these couplings cannot be physical free parameters anymore.

\subsection{Inclusion of \cred{a} fermion}
Now let us extend the theory space to $\Set{g_i}_{i=1,\dots,6}=\Set{ \xi_0, \lambda_0, \xi_2, \lambda_2, \lambda_4, y}$ \cred{with $N_\text{f}=1$}.
Solving the coupled equation $\beta_{g_i}=0$ with Eqs.~\eqref{xi0 RGE}--\eqref{lambda4 RGE}, we again obtain the Gaussian-matter fixed point:
\al{
\xi_0^*&=1.63\times 10^{-2},& \lambda_0^*&=3.72\times 10^{-3},& \xi_2^*&=0,& \lambda_2^*&=0,& \lambda_4^*&=0, & y^*&=0.
	\label{Gaussian-matter fixed point with a fermion}
}
Around this fixed point, the matrix~\eqref{beta matrix} becomes
\al{
M=
\pmat{
3.6814	&	-5.39674	&	0.00776027	&	-0.00258676	&		0		&	0\\
1.99718	& 	-0.663341	&	0.00295698	&	0.00534691	&		0		&	0\\
0		&	0		&	1.6814		&	-5.39674		&	-0.0155205	&	0\\
0		&	0		&	1.99718		&	-2.66334		&	0.0320815	&	0\\
0		&	0		&	0			&	0			&	-2.60696		&	0\\
0		&	0		&	0			&	0			&	0			&	-1.46426
}.
\label{matrixwithyukawa}
}
The eigenvalues $\theta_i$ and eigenvectors $V^{(i)}$ of the matrix \eqref{matrixwithyukawa} are
\al{
\theta_{1,2}&= 1.50903 \pm 2.46151i, &
V^{(1)},V^{(2)}&=
\pmat{
0.854336\\
0.343899\pm 0.389672 i\\
0\\
0\\
0\\
0
},\\
\theta_{3,4}&= -0.490968 \pm 2.46151i, &
V^{(3)},V^{(4)}&=
\pmat{
(-3.11425  \mp 1.27885i) \times 10^{-3}\\\
( -1.92736 \pm 0.618506i) \times 10^{-3}\\
0.854329\\
0.343897 \mp 0.389669i\\
0\\
0
},\\
\theta_5&= -2.60696, &
V^{(5)}&=
\pmat{
3.88533 \times 10^{-5}\\
2.91974 \times 10^{-5}\\
-1.65107 \times 10^{-2}\\
-1.59949 \times 10^{-2}\\
0.999736\\
0
},\\
\theta_6&= -1.46426, &
V^{(6)}&=
\pmat{
0\\
0\\
0\\
0\\
0\\
1
}.
}
Several comments are in order:
\begin{itemize}
\item We see from Eq.~\eqref{matrixwithyukawa} that there is no mixing between the Yukawa coupling and the others at this fixed point. As the critical exponent of $y$ is negative, the Yukawa interaction is irrelevant up to this truncation.
\item The critical exponents $\re\theta_{1,2}$ for the gravitational constants, namely the Newton constant $g_1$ ($=\xi_0=1/16\pi\tilde G$) and the cosmological constant $g_2$ ($=\lambda_0$), are substantially reduced from those in Sec.~\ref{scalar-gravity model} by the inclusion of fermions, even without the Yukawa coupling. This is due to the last terms in Eqs.~\eqref{xi0 RGE} and \eqref{lambda0 RGE}.
\item We see from Eqs.~\eqref{xi2 RGE}--\eqref{lambda4 RGE} that when $y=0$, the RG equations of 
the non-minimal coupling $g_3$ ($=\xi_2$),
the scalar mass-squared $g_4$ ($=\lambda_2$),
and the scalar quartic coupling $g_5$ ($=\lambda_4$)
do not differ from those in the scalar gravity model in Sec~\ref{scalar-gravity model}.
However, even without the Yukawa coupling, the fermion loops do affect the gravitational constants $g_{1,2}$ as above.
As a result, the critical exponents $\re\theta_{3,4}$ of $g_{3,4}$ turn to negative from the positive values in the scalar-gravity model.
The non-minimal coupling $g_3$ and the scalar mass-squared $g_4$ are both made irrelevant. These coupling constants are not on the UV critical surface anymore.
\end{itemize}

\cred{
On the last point, we note that the matter couplings $\lambda_2$, $\lambda_4$, $\xi_2$, and $y$ vanish at the Gaussian-matter fixed point and hence that they do not affect the critical exponents $\re\theta_{3,4}$ of the non-minimal coupling $g_3$ ($=\xi_2$) and the mass-squared $g_4$ ($=\lambda_2$). What is important for flipping the sign of the critical exponents is the fact that the fixed-point values for the gravitational sector, $g_1^*$ ($=\xi_0^*$) and $g_2^*$ ($=\lambda_0^*$), are reduced by the fermion loops.
Indeed, even if we put $N_\text{f}=0$ with the values~\eqref{Gaussian-matter fixed point with a fermion}, we still get $\re\theta_{3,4}=-0.508$. Also, even if we put $N_\text{f}=1$ for the values~\eqref{Gaussian-matter fixed point} without fermion loop, we still obtain $\re\theta_{3,4}=0.144$, which is very close to the true value 0.143 for $N_\text{f}=0$. Finally for illustration, we show analytic formulae for the submatrix of~$M$ in Eq.~\eqref{matrixwithyukawa}:
\al{
M_{33}
	&=	{51\paren{77-8N_\text{f}}\over\paren{1152\pi^2\xi_0^*-17}^2}
		-{27648\pi^2\paren{13824\pi^2\xi_0^*-104N_\text{f}+797}\over\paren{1152\pi^2\xi_0^*-17}^3}\lambda_0^*+\mathcal O\fn{\lambda_0^{*2}},\\
M_{34}
	&=	-{12\paren{27648\pi^2\xi_0^*-104N_\text{f}+593}\over\paren{1152\pi^2\xi_0^*-17}^2}\nn
	&\quad
		+{96\paren{576\pi^2\xi_0^*\paren{19584\pi^2\xi_0^*+739-72N_\text{f}}+4275-740N_\text{f}}\over\xi_0^*\paren{1152\pi^2\xi_0^*-17}^3}\lambda_0^*
		+\mathcal O\fn{\lambda_0^{*2}},\\
M_{43}
	&=	{180\paren{77-8N_\text{f}}\over\paren{1152\pi^2\xi_0^*-17}^2}\nn
	&\quad
		-{17\paren{-9667+1672N_\text{f}}+3456\pi^2\xi_0^*\paren{9667-1672N_\text{f}+4608\pi^2\xi_0^*\paren{93-4N_\text{f}+576\pi^2\xi_0^*}}\over
		32\pi^2\xi_0^{*2}\paren{1152\pi^2\xi_0^*-17}^3}\lambda_0^*
		+\mathcal O\fn{\lambda_0^{*2}},\\
M_{44}
	&=	{-9667+1672N_\text{f}+64\pi^2\xi_0^*\paren{-8837+432N_\text{f}+2304\pi^2\xi_0^*\paren{576\pi^2\xi_0^*-71}}\over32\pi^2\xi_0^{*2}\paren{1152\pi^2\xi_0^*-17}^2}\nn
	&\quad
		+{258309-44920N_\text{f}-9216\pi^2\xi_0^*\paren{-3641+414N_\text{f}-144\pi^2\xi_0^*\paren{6912\pi^2\xi_0^*+893-24N_\text{f}}}\over16\pi^2\xi_0^*\paren{1152\pi^2\xi_0^*-17}^3}\lambda_0^*
		+\mathcal O\fn{\lambda_0^{*2}}.
}
}

\section{Summary and discussions}\label{summary}
In this paper we have investigated the fixed point structure of the Higgs-Yukawa model that is non-minimally coupled to gravity, using the FRG.
The full set of RG equations of this system are obtained for the first time.
We find a Gaussian-matter fixed point which is a non-trivial UV fixed point for the gravitational coupling constants, namely the Newton and cosmological constants, and is a trivial one for the other  coupling constants among matters. 
It has been known that the Gaussian-matter fixed point for the scalar-gravity system without fermion has the non-minimal coupling $\xi_2$ as relevant direction, together with the scalar mass-squared ($=\lambda_2$)~\cite{Percacci:2003jz,Narain:2009fy,Narain:2009gb}.
We have found that the inclusion of fermion to this scalar-gravity system makes both of them irrelevant, no matter whether the Yukawa coupling is turned on or not.
Therefore both of them in this toy model cannot be on the UV critical surface, and hence cannot be the free parameters of the theory in the asymptotic safety scenario.

It is important to investigate whether the non-minimal coupling of the Higgs to the Ricci scalar in the SM (and its extensions), $\xi\ab{H}^2R$, becomes relevant or not when we take into account the large degrees of freedom, both bosonic and fermionic, that couple to the Higgs.
The large non-minimal coupling constant plays crucial role in the Higgs inflation scenario~\cite{Bezrukov:2007ep,Hamada:2014iga}.
If the non-minimal coupling becomes a free parameter in the asymptotic safety scenario in the above sense, then we can use it to account for the cosmological data by the Higgs inflation.
In this toy model, we have found that the non-minimal coupling cannot be such a free parameter.
If this is the case for the SM too, then the Higgs inflation model is a cutoff theory and cannot be a UV complete model within the asymptotically safe gravity scenario.

In this paper, we have have studied the asymptotic safety of the simple Higgs-Yukawa model with non-minimal coupling, in the symmetric phase $\langle\phi\rangle=0$.
In the Higgs inflation using the SM criticality, the typical value of the Higgs field  becomes close to the Planck scale~\cite{Hamada:2014wna}.
For such an application, it is important to extend our analysis to the broken phase $\langle \phi \rangle \neq 0$.

We comment on the unitarity of gravity.
The earlier studies indicate that the asymptotically safe quantum gravity would be described by the three dimensional UV critical surface, spanned by the cosmological constant, $R$, and $R^2$; see e.g.~\cite{Reuter:2012id,Falls:2014tra}.
It is worth studying whether this remains the case or not if we include other forms of higher dimensional operators.
For example, the operators $R_{\mu\nu}R^{\mu\nu}$ and $R_{\mu\nu\rho\sigma}R^{\mu\nu\rho\sigma}$ have not been taken into account in the literature although they give the same order of contribution as $R^2$, due to technical difficulties in distinguishing these three in the heat kernel expansions around the $S^4$ background; see e.g.\ Refs.~\cite{Lauscher:2001rz,Lauscher:2001ya,Lauscher:2002sq}.
It is very important to include these terms beyond the current truncation.
If they turn out to take part in the UV critical surface then the UV gravity is not unitary anymore in general.\footnote{
The standard line of reasoning that higher-derivative gravity leads to ghost poles in the propagator and thus violates unitarity is not necessarily applicable in the asymptotic safety context: Towards the UV, the FRG propagators are still regularized and thus there are no ghost poles by construction, though this does not mean that the theory is automatically unitary. The issue of unitarity can only be clarified once all fluctuations are integrated out and the resulting vacuum state turns out to be stable (with Minkowski signature). These aspects are discussed in more detail in Ref.~\cite{Niedermaier:2006wt}. We thank H.\ Gies for clarifying this point.
}
It might also be interesting if the theory still remains meaningful under such a situation.

In this paper, we have limited ourselves within the LPA where  we neglect the field renormalization and see only the local couplings without external momenta.
LPA has been a useful tool to investigate e.g.\ the vacuum structure of the quantum chromodynamics.
Although we expect that the LPA is applicable for sufficiently homogeneous field configurations, it is not clear how good an approximation the LPA is for the analysis of the asymptotically safe gravity.
It would be useful to go beyond the LPA by taking into account the anomalous dimensions from the field renormalization.

If the quartic scalar coupling $\lambda_4$ has a non-trivial UV fixed point, namely, $\lambda_4^*\neq0$ and $\lambda_4\phi^4$ is relevant around the fixed point, then it becomes a solution to the triviality of the scalar $\phi^4$ theory; see e.g.~\cite{Gies:2013pma,Gies:2015lia}. 
It would be interesting to study the triviality under the presence of gravity extending the study in Refs.~\cite{Vacca:2010mj}, by including e.g.\ the non-minimal coupling.

We comment on the so-called hierarchy problem of the Higgs mass-squared.
Let us write the dimensionful mass-squared $m^2\fn{\Lambda}:=2\lambda_2\fn{\Lambda}\Lambda^2$ at the scale $\Lambda$. This reduces to the bare mass at the UV cutoff scale: $m^2\fn{\Lambda_0}=m_0^2$. For illustration purpose, let us switch off all the coupling constants except for $\lambda_2$ and $y$ in the RG equation for the mass-squared~\eqref{lambda2 RGE}, namely, we take the limits $\lambda_0\to0$, $\xi_0={1\over16\pi\tilde G}\to\infty$, $\xi_2\to 0$, and $\lambda_4\to0$: 
\al{
-\Lambda{\p\over\p\Lambda}m^2
	&=	-{N_\text{f}y^2\over4\pi^2}\Lambda^2.
		\label{toy RGE}
}
If we neglect the running of $y$, we get
\al{
m^2\fn{\Lambda^2}
	&\simeq
		m_0^2-{N_\text{f}y^2\over8\pi^2}\Lambda_0^2+{N_\text{f}y^2\over8\pi^2}\Lambda^2.
			\label{toy Higgs mass}
}
At very low scales $\Lambda\ll\Lambda_0$, the mass-sqaured becomes
$m^2\fn{\Lambda^2}\to m_0^2-{N_\text{f}y^2\over8\pi^2}\Lambda_0^2$
and we need the fine-tuning between the bare mass-squared and the loop correction if we want $m^2\fn{\Lambda^2}\ll\Lambda_0^2$ ($\sim m_0^2$).
This is the fine-tuning problem.

This problem still remains in the SM, in principle, even under the asymptotically safe gravity, e.g.\  considered in Ref~\cite{Shaposhnikov:2009pv}:
Suppose we start from the UV cutoff scale much larger than the Planck scale, $\Lambda_0\gg1/\sqrt{32\pi G}$.
Naively, if the dimensionless mass-squared $\lambda_2\fn{\Lambda}$ turns to be irrelevant around the UV fixed point as in our result, one might expect that it could be a solution to the hierarchy problem.
However, even if we start from small $\lambda_2\fn{\Lambda}$ near the UV fixed point $\Lambda\sim\Lambda_0$, and further gets the exponential suppression due to its irrelevance in the RG evolution  departing from the UV fixed point along the UV critical surface, eventually $\lambda_2\fn{\Lambda}$ will mix with other relevant operators in the coupled non-linear evolution down to the Planck scale, and the resultant mass will be of the order of the Planck scale in general, $\lambda_2\fn{\Lambda}\Lambda^2\sim1/32\pi G$. It would be interesting to look for a mechanism to keep $\lambda_2\fn{\Lambda}$ tiny for the scales down to the Planck scale.
Then this sets the boundary condition at the Planck scale for the subsequent RG evolution further down to the electroweak scale, in which
$\lambda_2\fn{\Lambda}\Lambda^2$ and $1/\sqrt{32\pi G}$ correspond to $m_0^2$ and $\Lambda_0$ in Eq.~\eqref{toy Higgs mass}, respectively. If we further manage to find a mechanism to make the sum of SM loop corrections, corresponding to the second term in the right hand side of Eq.~\eqref{toy Higgs mass}, to vanish, as is speculated by Veltman~\cite{Veltman:1980mj}, then the fine-tuning problem is solved. Note that the observed Higgs mass allows the Veltman condition to be satisfied at the Planck scale and that the two loop correction to the Veltman condition is negligibly small~\cite{Hamada:2012bp}, although the theoretical explanation why it holds is still missing.
Similarly the cosmological constant problem is yet to be solved.

\subsection*{Acknowledgement}
We thank Ken-Ichi Aoki, Holger Gies, Yuta Hamada, \cred{Aaron Held,} Hikaru Kawai, \cred{Jan M.\ Pawlowski,} Roberto Percacci, Gian Paulo Vacca, Satoshi Yamaguchi, and Ryo Yamamura for useful discussions and comments.
\cred{We are grateful to the referees for careful reading and providing the constructive comments.}
M.Y.\ thanks the hospitality of the Particle Physics Theory Group at Osaka University during his stays.
We thank Yuyake Jet at which this collaboration has started.
The work of K.O.\ is in part supported by the Grant-in-Aid for Scientific Research Nos.~23104009 and 15K05053.
The work of M.Y.\ is supported by a Grant-in-Aid for JSPS Fellows (No. 25-5332).

\appendix
\section*{Appendix}
\section{Wetterich equation}\label{FRGnote}
We briefly sketch out the derivation of the Wetterich equation using a simple scalar theory in flat spacetime without employing the background field method.
Physically, we will derive the effective action $\Gamma_\Lambda$ with the cutoff $\Lambda$, from the bare action $S_0$ defined at the UV cutoff scale~$\Lambda_0$.
Note that in the asymptotic safety scenario, the UV finite theory is defined on the finite dimensional UV critical surface that consists of the renormalized trajectories flowing out of the UV fixed point: We define the bare theory at a point on one of such renormalized trajectories. The choice of this point and the scale $\Lambda_0$ assigned to it are more or less arbitrary, given the point is right on the renormalized trajectory.\footnote{
The asymptotic safety is somewhat contrary to the ordinary low-energy effective field theory picture in the sense that we must define the theory right on the UV critical surface and that even an infinitesimal displacement from it results in the divergence from it when we track back the renormalization flow toward UV direction. That is, if we write down all the possible operators at IR scales allowed by symmetry, then there is infinitesimally small chance to reach the asymptotically safe theory when we trace back the renormalization group flow toward UV direction.
}

We write the partition function 
\begin{align}
Z_\Lambda[J]
	=	e^{W_\Lambda[J]}
	:=	\int\sqbr{\mathcal D\varphi}e^{-S_0[\varphi]-\Delta S_\Lambda[\varphi]+ J\cdot\varphi}
			\label{partition function}
\end{align} 
with the regulator term
\al{
\Delta S_\Lambda[\varphi]
	&:=	{1\over2}\int\frac{\df^4p}{(2\pi) ^4}\varphi\fn{-p}^\T R_\Lambda\fn{p} \varphi\fn{p},
	\label{regulator term}
}
where $W_\Lambda[J]$ is the generating functional of connected diagrams;
we write $\int_x:=\int\df^4x$;
and we introduce the cutoff profile function
\al{
R_\Lambda\fn{p}
	&\sim	\begin{cases}
				\Lambda^2	&	\text{for $p<\Lambda$},\\
				0			&	\text{for $p>\Lambda$},
			\end{cases}
}
which suppresses the lower momentum modes with $p<\Lambda$ and leaves the higher ones with $\Lambda < p <\Lambda_0$. 
That is, the low momentum modes with $p<\Lambda$ are given the extra mass $\Lambda$ in the path integral~\eqref{partition function} and are not effectively path-integrated in the partition function~\eqref{partition function}.
Therefore, $\Lambda$ can be interpreted as a new UV cutoff scale in $Z_\Lambda[J]$ in which the high momentum modes with $\Lambda<p<\Lambda_0$ are integrated out.
Namely, $\Lambda$ is the IR cutoff scale for the integrated high momentum modes, and the UV cutoff scale for the unintegrated low momentum modes.
We also introduce the position-space cutoff function $R_\Lambda\fn{x,y}=R_\Lambda\fn{y,x}$ by
\al{
\Delta S_\Lambda[\varphi]
	=	{1\over2}\int_x\int_y\,\varphi\fn{x} ^\T R_\Lambda\fn{x,y} \varphi\fn{y}.
}

The effective action $\Gamma_\Lambda$ is given by the Legendre transformation of $W_\Lambda$:
\al{
\Gamma_\Lambda[\Phi]
	&:=	J^\Phi_\Lambda\cdot\Phi -W_\Lambda{[J^\Phi_\Lambda]} -\Delta S_\Lambda{[\Phi]},
}
where $J^\Phi_\Lambda$ is defined by
\al{
{\delta W_\Lambda\over\delta J\fn{x}}[J^\Phi_\Lambda]
	&=	\Phi\fn{x}.
}
Then we get
\al{
{\delta\Gamma_\Lambda\over\delta\Phi\fn{x}}[\Phi]
	&=	\int_y{\delta J^\Phi_\Lambda\fn{y}\over\delta\Phi\fn{x}}\Phi\fn{y}
		+J^\Phi_\Lambda\fn{x}
		-\int_y{\delta W_\Lambda\over\delta J\fn{y}}[J^\Phi_\Lambda]\,
			{\delta J^\Phi_\Lambda\fn{y}\over\delta\Phi\fn{x}}
		-{\delta\Delta S_\Lambda\over\delta\Phi\fn{x}}[\Phi]\nn
	&=	J^\Phi_\Lambda\fn{x}-\int_yR_\Lambda\fn{x,y}\Phi\fn{y},
}
and further
\al{
{\delta^2\Gamma_\Lambda\over\delta\Phi\fn{x}\delta\Phi\fn{y}}[\Phi]
	&=	{\delta J^\Phi_\Lambda\fn{x}\over\delta\Phi\fn{y}}-R_\Lambda\fn{x,y}.
			\label{d2Gamma/dPhi2}
}
We similarly define $\Phi^J_\Lambda$ by
\al{
{\delta W_\Lambda\over\delta J\fn{x}}[J]
	&=	\Phi^J_\Lambda\fn{x}
		\label{dW/dJ}
}
so that $\Phi^J_\Lambda=\Phi$ if $J=J^\Phi_\Lambda$.
Taking a functional derivative of Eq.~\eqref{dW/dJ}, we get
\al{
{\delta^2 W_\Lambda\over\delta J\fn{x}\delta J\fn{y}}[J]
	&=	{\delta\Phi^J_\Lambda\fn{x}\over\delta J\fn{y}}[J],
}
and hence
\al{
{\delta^2 W_\Lambda\over\delta J\fn{x}\delta J\fn{y}}[J_\Lambda^\Phi]
	&=	{\delta\Phi^J_\Lambda\fn{x}\over\delta J\fn{y}}[J_\Phi^\Lambda]
	=	\paren{{\delta J^\Phi_\Lambda\fn{x}\over\delta \Phi\fn{y}}[\Phi]}^{-1}
	=	\paren{{\delta^2\Gamma_\Lambda\over\delta\Phi\fn{x}\delta\Phi\fn{y}}[\Phi]+R_\Lambda\fn{x,y}}^{-1},
		\label{d2W/dJ2}
}
where the inverse is in the functional space spanned by $x$ and $y$ and we have used Eq.~\eqref{d2Gamma/dPhi2} in the last step.

We want to evaluate
\al{
{\df\Gamma_\Lambda[\Phi]\over\df\Lambda}
	&=	\int_x{\df J^\Phi_\Lambda\fn{x}\over\df\Lambda}\Phi\fn{x}
		-{\df W_\Lambda[J^\Phi_\Lambda]\over\df\Lambda}
		-{\df\Delta S_\Lambda[\Phi]\over\df\Lambda}.
		\label{Gamma derivative}
}
After some computation, the second term in Eq.~\eqref{Gamma derivative} becomes\footnote{
Concretely,
\als{
-{\df W_\Lambda[J^\Phi_\Lambda]\over\df\Lambda}
	&=	{1\over Z[J^\Phi_\Lambda]}\int\sqbr{\mathcal D\varphi}
			e^{-S_{\Lambda_0}[\varphi]-\Delta S_\Lambda[\varphi]+ J^\Phi_\Lambda\cdot\varphi}
			\paren{
				{\df\Delta S_\Lambda[\varphi]\over\df\Lambda}
				-{\df J^\Phi_\Lambda\over\df\Lambda}\cdot\varphi
				}\nn
	&=	{1\over Z[J^\Phi_\Lambda]}\int\sqbr{\mathcal D\varphi}
			e^{-S_{\Lambda_0}[\varphi]-\Delta S_\Lambda[\varphi]+ J^\Phi_\Lambda\cdot\varphi}
			\paren{
				\int_x\int_y{1\over2}\varphi\fn{x}{\df R_\Lambda\fn{x,y}\over\df\Lambda}\varphi\fn{y}
				-\int_x{\df J^\Phi_\Lambda\fn{x}\over\df\Lambda}\varphi\fn{x}
				}\nn
	&=	\paren{{1\over Z[J^\Phi_\Lambda]}\int\sqbr{\mathcal D\varphi}
			e^{-S_{\Lambda_0}[\varphi]-\Delta S_\Lambda[\varphi]+ J^\Phi_\Lambda\cdot\varphi}
				\int_x\int_y{1\over2}\varphi\fn{x}{\df R_\Lambda\fn{x,y}\over\df\Lambda}\varphi\fn{y}}
			-\int_x{\df J^\Phi_\Lambda\fn{x}\over\df\Lambda}{\delta W_\Lambda\over\delta J\fn{x}}[J^\Phi_\Lambda]\nn
	&=	\paren{{1\over Z[J^\Phi_\Lambda]}\int\sqbr{\mathcal D\varphi}
				e^{-S_{\Lambda_0}[\varphi]-\Delta S_\Lambda[\varphi]+ J^\Phi_\Lambda\cdot\varphi}\,
				{1\over2}\int_x\int_y\varphi\fn{x}{\df R_\Lambda\fn{x,y}\over\df\Lambda}\varphi\fn{y}}
			-\int_x{\df J^\Phi_\Lambda\fn{x}\over\df\Lambda}\Phi\fn{x}\nn
	&=	{1\over2Z[J^\Phi_\Lambda]}\int_x\int_y
			{\delta^2Z\over\delta J\fn{x}\delta J\fn{y}}[J^\Phi_\Lambda]\,
			{\df R_\Lambda\fn{x,y}\over\df\Lambda}
		-\int_x{\df J^\Phi_\Lambda\fn{x}\over\df\Lambda}\Phi\fn{x}\nn
	&=	{1\over2}\int_x\int_y
			\paren{
				{\delta^2W_\Lambda\over\delta J\fn{x}\delta J\fn{y}}[J^\Phi_\Lambda]
				+\Phi\fn{x}\Phi\fn{y}
				}
			{\df R_\Lambda\fn{x,y}\over\df\Lambda}
		-\int_x{\df J^\Phi_\Lambda\fn{x}\over\df\Lambda}\Phi\fn{x}\nn
	&=	{1\over2}\int_x\int_y
				{\delta^2W_\Lambda\over\delta J\fn{x}\delta J\fn{y}}[J^\Phi_\Lambda]\,
				{\df R_\Lambda\fn{x,y}\over\df\Lambda}
		+{\df\Delta S_\Lambda[\Phi]\over\df\Lambda}
		-\int_x{\df J^\Phi_\Lambda\fn{x}\over\df\Lambda}\Phi\fn{x},
}
where we have used
\als{
{\delta W_\Lambda\over\delta J\fn{x}}[J]
	&=	{1\over Z[J]}{\delta Z_\Lambda\over\delta J\fn{x}}[J],	&
{\delta^2 W_\Lambda\over\delta J\fn{x}\delta J\fn{y}}[J]
	&=	{1\over Z[J]}{\delta^2 Z_\Lambda\over\delta J\fn{x}\delta J\fn{y}}[J]
		-\Phi^J_\Lambda\fn{x}\Phi^J_\Lambda\fn{y}.
}
}
\al{
-{\df W_\Lambda[J^\Phi_\Lambda]\over\df\Lambda}
	&=	{1\over2}\int_x\int_y
				{\delta^2W_\Lambda\over\delta J\fn{x}\delta J\fn{y}}[J^\Phi_\Lambda]\,
				{\df R_\Lambda\fn{x,y}\over\df\Lambda}
		+{\df\Delta S_\Lambda[\Phi]\over\df\Lambda}
		-\int_x{\df J^\Phi_\Lambda\fn{x}\over\df\Lambda}\Phi\fn{x}.
}
Therefore,
\al{
{\df\Gamma_\Lambda[\Phi]\over\df\Lambda}
	&=	{1\over2}\int_x\int_y
				{\delta^2W_\Lambda\over\delta J\fn{x}\delta J\fn{y}}[J^\Phi_\Lambda]\,
				{\df R_\Lambda\fn{x,y}\over\df\Lambda}.
					\label{dGamma/dLambda}
}
Putting Eq.~\eqref{d2W/dJ2} into Eq.~\eqref{dGamma/dLambda}, we get the Wetterich equation
\al{
{\df\Gamma_\Lambda[\Phi]\over\df\Lambda}
	&=	{1\over2}\int_x\int_y
				\paren{
					{\delta^2\Gamma_\Lambda\over\delta\Phi\fn{x}\delta\Phi\fn{y}}[\Phi]
					+R_\Lambda\fn{x,y}
					}^{-1}\,
				{\df R_\Lambda\fn{x,y}\over\df\Lambda}.
}
For general case including fermions, this expression becomes Eq.~\eqref{wetterich}.
We see that the Wetterich equation is one-loop exact from its derivation.

\section{Supertrace}\label{supermatrix}
A supertrace of a supermatrix
\al{
M
	&=	\bmat{
			M_\BB	&	M_\BF\\
			M_\FB	&	M_\FF
			}
}
is defined by
\al{
\str M
	&=	\tr M_\BB-\tr M_\FF,
}
which satisfies $\str\fn{ M N}=\str\fn{ N M}$.
A superdeterminant is defined by
\al{
\sdet M
	&=	\exp\fn{\str\ln M},
}
satisfying
\al{
\sdet\fn{ M N}
	&=	\sdet M\,\sdet N.
}
We see
\al{
\sdet M
	&=	\sdet\bmat{M_\BB&0\\ M_\FB&1}\,
		\sdet\bmat{
				1&M_\BB^{-1}M_\BF\\
				0&M_\FF-M_\FB M_\BB^{-1}M_\BF
				}\nn
	&=	{\det M_\BB\over\det\fn{M_\FF-M_\FB M_\BB^{-1}M_\BF}}.\label{smat1}
}
If we decompose the matrix as
\al{
M = 
\bmat{
M_\BB -M_\BF M_\FF^{\cblue{-1}} M_\FB	&	M_\BF M_\FF ^{-1} \\
0	&	1
}
\bmat{
1	&	0\\
M_\FB	&	M_\FF
},
}
we see
\al{
\sdet M
	&=	{\det \fn{M_\BB -M_\BF M_\FF^{\cblue{-1}} M_\FB }\over \det M_\FF}.
}
In this paper we use \eqref{smat1}.

\section{Functional renormalization group for the effective action}\label{frgtreatment}
In Appendix~\ref{FRGnote}, we have briefly reviewed the derivation of the Wetterich equation for a simple scalar case.
For general case including fermions, the Wetterich equation reads~\cite{Wetterich:1992yh,Morris:1993qb}
\begin{align}
{\p\over\p\Lambda} \Gamma_\Lambda
	&=	\frac{1}{2}\STr_{x,y}\left[ \left( \frac{\overrightarrow \delta}{\delta \Phi\fn{x}} \Gamma _\Lambda\frac{\ola \delta}{\delta \Phi\fn{y}}+\mc R_\Lambda\fn{x,y}  \right) ^{-1}\cdot {\p\over\p\Lambda} \mc R_\Lambda\fn{y,x}\right].
	\label{wetterich}
\end{align}
For later convenience, let us briefly review how to treat the supermatrix in the Wetterich equation.

We separate the two-point function and the cutoff function $\mc R_\Lambda$ into bosonic and fermionic parts, respectively,
\al{
\bmat{
\Gamma_\BB &\Gamma_\BF\\
\Gamma_\FB &\Gamma_\FF
}
	&:=	\frac{\overrightarrow \delta}{\delta \Phi\fn{x}} \Gamma _\Lambda\frac{\ola \delta}{\delta \Phi\fn{y}},	&
\bmat{
\mc R_\BB &0\\
0 &\mc R_\FF
}
	&:=	\mc R_\Lambda\fn{x,y}.
}
We also define
\al{
\bmat{
{\mc M}_\BB &{\mc M}_\BF\\
{\mc M}_\FB &{\mc M}_\FF
}
	&:=	\bmat{
			\Gamma_\BB &\Gamma_\BF\\
			\Gamma_\FB &\Gamma_\FF
			}
		+
		\bmat{
			\mc R_\BB &0\\
			0 &\mc R_\FF
			}.
			\label{curly M}
}
Then we rewrite the Wetterich equation as
\al{
{\p\over\p\Lambda}\Gamma_\Lambda
	&=	\wt{\p\over\p\Lambda}{1\over2}\ln\fn{\SDet\bmat{
			{\mc M}_\BB &{\mc M}_\BF\\
			{\mc M}_\FB &{\mc M}_\FF
			}}\nn
	&=	{1\over2}\wt{\p\over\p\Lambda}\paren{\ln\bigg[\Det  {\mc M}_\BB\bigg]
			-\ln\bigg[\Det\Big({\mc M}_\FF-{\mc M}_\FB {\mc M}_\BB^{-1}{\mc M}_\BF\Big)\bigg]
			}\nn
	&=	{1\over2}\wt{\p\over\p\Lambda}\Tr\bigg[\ln {\mc M}_\BB\bigg]
			-{1\over2}\wt{\p\over\p\Lambda}\Tr\bigg[\ln\Big({\mc M}_\FF-{\mc M}_\FB {\mc M}_\BB^{-1}{\mc M}_\BF\Big)\bigg],
				\label{preWetterich}
}
where $\wt{\p/\p\Lambda}$ acts only on $\mc R_\BB$ and $\mc R_\FF$ and we used the formulation for supermatrix summarized in Appendix~\ref{supermatrix}.
Performing the derivative, we obtain
\al{
{\p\over\p\Lambda}\Gamma_\Lambda
	&=	{1\over2}\Tr\bigg[{\mc M}_\BB^{-1}{\p \mc R_\BB\over\p\Lambda}\bigg]\nn
	&\quad
		-{1\over2}\Tr\bigg[\Big({\mc M}_\FF-{\mc M}_\FB {\mc M}_\BB^{-1}{\mc M}_\BF\Big)^{-1}
					\paren{{\p \mc R_\FF\over\p\Lambda}+{\mc M}_\FB {\mc M}_\BB^{-1}\,
						{\p \mc R_\BB\over\p\Lambda}\,
						{\mc M}_\BB^{-1}{\mc M}_\BF}\bigg].
						\label{FRG_total}
}
The first term in RHS of \eqref{FRG_total} is the fluctuations of bosonic fields.
The second term includes not only the fluctuations of fermionic fields but also the mixing of fermion and boson. We have directly computed the algebraic expression in the right hand side of Eq.~\eqref{FRG_total} to cross-check the results in Sec.~\ref{models} that is obtained diagrammatically.

We may also expand the second term in Eq.~\eqref{preWetterich} as
\al{
-{1\over2}\wt{\p\over\p\Lambda}\Tr\bigg[\ln\Big({\mc M}_\FF-{\mc M}_\FB {\mc M}_\BB^{-1}{\mc M}_\BF\Big)\bigg]
	&=	-{1\over2}\wt{\p\over\p\Lambda}\Tr\bigg[\ln {\mc M}_\FF+\ln\Big(1-{\mc M}_\FF^{-1}{\mc M}_\FB {\mc M}_\BB^{-1}{\mc M}_\BF\Big)\bigg]\nn
	&=	-{1\over2}\wt{\p\over\p\Lambda}\paren{
			\Tr\bigg[\ln {\mc M}_\FF\bigg]
			-\Tr\bigg[{\mc M}_\FF^{-1}{\mc M}_\FB {\mc M}_\BB^{-1}{\mc M}_\BF\bigg]+\cdots
			}\nn
	&=	-{1\over2}\Tr\bigg[{\mc M}_\FF^{-1}\,{\p \mc R_\FF\over\p\Lambda}\bigg]\nn
	&\quad
		-{1\over2}\Tr\bigg[{\mc M}_\FF^{-1}\,{\p \mc R_\FF\over\p\Lambda}\,{\mc M}_\FF^{-1}{\mc M}_\FB {\mc M}_\BB^{-1}{\mc M}_\BF\bigg]\nn
	&\quad
		-{1\over2}\Tr\bigg[{\mc M}_\FF^{-1}{\mc M}_\FB {\mc M}_\BB^{-1}\,{\p \mc R_\BB\over\p\Lambda}\,{\mc M}_\BB^{-1}{\mc M}_\BF\bigg]
		+\cdots\cred{,} \label{frgequation}
}
\cred{
where the higher order terms, represented by dots, are all higher-dimensional operators being already truncated in Eq.~\eqref{effectiveaction}, and hence we neglect them in this paper.
}
This expression~\eqref{frgequation} is useful to compare with the Feynman diagramatic computation since the vertex structure is clearer.
It is especially useful when evaluating the beta function of the Yukawa coupling constant.
We have also used this expression to further cross-check the results in Sec.~\ref{models}.

\section{Heat kernel trace}\label{hkeapp}
In this section we briefly review how to take the trace in the heat kernel expansion; see e.g.\ Refs.~\cite{Codello:2008vh,Vassilevich:2003xt,Dunne:2007rt} for more detailed reviews.
Consider an arbitrary function $W\fn{p^2}$ and its trace $\Tr\sqbr{W\fn{p^2}}$.
Using the Laplace transformation, we get
\al{\label{hktr}
\Tr\sqbr{W\fn{p^2}} =\int ^\infty_{-\infty} \df s\, {\tildeForLaplace W}\fn{s}\Tr\sqbr{e^{-s\,p^2}}.
}
The trace in the right-hand side can be expanded as
\al{
\Tr\sqbr{e^{-s\,p^2}}	= \sum _{n=0}B_n\fn{p^2} s^{-\paren{D-n}/2},
}
where
\al{
\label{hke}
B_n\fn{p^2}=\frac{1}{(4\pi)^{D/2}}\int \df^Dx\sqrt{g}\, \tr\sqbr{{\bf b}_n}.
}
The heat kernel coefficients ${\bf b}_n$ are given by ${\bf b}_0 = {\bf 1}$, ${\bf b}_2 = {R\over 6}\,{\bf 1}$, etc., where ${\bf 1}$ is the identity on the spin representation of the field. Their explicit values are shown in Table~\ref{hkcs}. For higher order ($n>2$), see e.g., the appendix of \cite{Codello:2008vh}.
By inserting (\ref{hke}) into the right-hand side of (\ref{hktr}), we obtain
\al{\label{traceexplict}
\Tr \,[W(p^2)] = \frac{1}{(4\pi)^{D/2}}\bigg\{ Q_{\frac{D}{2}}[W] \int \df^Dx \sqrt{g} \tr[\bar {\bf b}_0]+ Q_{\frac{D}{2}-1}[W] \int \df^Dx\sqrt{g}R\tr[\bar {\bf b}_2] +{\mO}(R^2) \bigg\},
}
where $\bar {\bf b}_0:= {\bf 1}$, $\bar {\bf b}_2:={\bf 1}/6$, and
\al{
Q_n[W] := \int^\infty_{-\infty}\df s(-s)^{-n}{\tildeForLaplace W}(s).
}
Its Mellin transformation yields
\al{
Q_0[W] &= W(0), \nn
Q_n[W] &= \frac{1}{\Gamma[n]}\int^\infty_0\df z\, z^{n-1}W[z],
\label{mellinQ}
}
where $\Gamma[n]$ is the Gamma function. Thanks to above relations (\ref{traceexplict}) and (\ref{mellinQ}), the trace for the eigenvalues of the derivative operator can be evaluated in curved space.

\begin{table}[htb]
\begin{center}
\caption{heat kernel coefficients for the individual fields in $D=4$}
\label{hkcs}
\begin{tabular}{|c|c|c|c|c|}
\hline
 &　$h^\perp_{\mu\nu}$ (spin 2)  &  $\xi_\mu$, $C^\perp_\mu$ (spin 1)& $\psi, {\overline \psi}$ (spin 1/2)& $h, \sigma, \phi, C, {\bar C}$ (spin 0) \\
\hline
$\tr[\bar {\bf b}_0]=: b_0$ & $5$ & $3$ & 4 &  1 \\
\hline
$\tr[\bar {\bf b}_2]=:b_2$ & $\displaystyle -\frac{5}{6}$ & $\displaystyle \frac{1}{4}$ & $\displaystyle \frac{2}{3}$ & $\displaystyle \frac{1}{6}$ \\
\hline
\end{tabular}
\end{center}
\label{tbl:ltx-tbl2}
\end{table}
\section{Derivation of the beta functions}\label{explicitcal}

In this appendix we show the diagrammatic derivation of the beta functions.
\subsection{Bosonic contributions}
We evaluate the first term in Eq.~\eqref{FRG_total} corresponding to the boson loops:
\al{
{1\over2}\Tr\bigg[{\mc M}_\BB^{-1}{\p \mc R_\BB\over\p\Lambda}\bigg]=
	\frac{1}{2}{\rm Tr}\left. \frac{\p_t{\mathcal R}_\Lambda}{\Gamma _\Lambda^{(1,1)}	
					+{\mathcal R}_\Lambda}\right|_{h^\perp h^\perp}
							+\frac{1}{2}{\rm Tr}'\left. \frac{\p_t{\mathcal R}_\Lambda}{\Gamma_\Lambda^{(1,1)}
									+{\mathcal R}_\Lambda}\right|_{\xi \xi}
						+\frac{1}{2}{\rm Tr}''\left. \frac{\p_t{\mathcal R}_\Lambda}{\Gamma _\Lambda^{(1,1)}
									+{\mathcal R}_\Lambda}\right|_{\rm SS}.
				\label{scalar_loops}
}
We evaluate these contributions part by part using the explicit form of two-point functions and cutoff functions exhibited in subsection \ref{modelandmanipulations}.
\subsubsection{The loop contribution of the transverse gravity field}
The loop contribution of $h^\perp$ field is evaluated:
\al{
\frac{1}{2}{\rm Tr}\left. \frac{\p_t{\mathcal R}_\Lambda}{\Gamma _\Lambda^{(1,1)}	
					+{\mathcal R}_\Lambda}\right|_{h^\perp h^\perp}
				&=	\frac{1}{2}{\rm Tr}~\frac{\frac{1}{2}\paren{\p_t F}R_\Lambda	
				+\frac{1}{2}F\paren{\p_t R_\Lambda}}{\frac{1}{2}F(P_\Lambda+\frac{2}{3}{R})
					-\frac{1}{2}V-\frac{1}{2}Y}\nn
				&= \frac{1}{2}{\rm Tr}~\frac{\paren{\p_t F}R_\Lambda	+F\paren{\p_t R_\Lambda}}{FP_\Lambda-V-Y}
				- \frac{1}{3}{\rm Tr}~\frac{\paren{\p_t F}R_\Lambda	+F\paren{\p_t R_\Lambda}}{(FP_\Lambda-V-Y)^2}\, F{R}	+\mO ({R}^2).
}
Using the heat kernel expansion given in Appendix~\ref{hkeapp}, we evaluate the trace for $\mathcal O(R^0)$,
\al{
\frac{1}{2}{\rm Tr}[W_{h^\perp}^1] &:= \frac{1}{2}{\rm Tr}\, \frac{\paren{\p_t F}R_\Lambda	+F\paren{\p_t R_\Lambda}}{FP_\Lambda-V-Y}\nn
				&=	\frac{1}{2}\frac{1}{\paren{4\pi}^2}\left\{ b_0^{h^\perp} Q_2[W_{h^\perp}^1]\int \df^4x\sqrt{g}	
							+	b_2^{h^\perp} Q_1[W_{h^\perp}^1]\int \df^4x\sqrt{g}{R} \right\} +\mO({R}^2)
}
with the functions
\al{
\nonumber
Q_2[W_{h^\perp}^1]	&=		\frac{1}{\Gamma(2)}\int _0^{\infty}\df z\, zW_{h^\perp}^1(z)
			=	
				\int _0^{\infty}\df z\, z\frac{\paren{\p_tF}R_\Lambda		+F\paren{\p_tR_\Lambda}}{FP_\Lambda-V-Y}
\nn
			&=\frac{\Lambda^6}{F\Lambda^2-V}\left[ \frac{1}{6}\paren{\p_tF}-F\right]
			+\frac{\Lambda^6}{\paren{F\Lambda^2-V}^2}\left[ \frac{1}{6}\paren{\p_tF}-F\right]\, Y + \mO(Y^2),
\\
Q_1[W_{h^\perp}^1]	&=	\frac{1}{\Gamma(1)}\int ^{\infty}_0\df z~W_{T1}(z)
			=	
				\int ^{\infty}_0\df z~\frac{\paren{\p_tF}R_\Lambda+F\paren{\p_tR_\Lambda}}{FP_\Lambda-V-Y}\nn
			&=\frac{\Lambda^4}{F\Lambda^2-V}\left[ \frac{1}{2}\paren{\p_tF}-2F\right] +\mO(Y),
}
and for $\mathcal O(R)$,
\al{
-\frac{1}{3}{\rm Tr}[W_{h^\perp}^2] {R} &:=- \frac{1}{3}{\rm Tr}\, \frac{\paren{\p_t F}R_\Lambda	+F\paren{\p_t R_\Lambda}}{(FP_\Lambda-V-Y)^2}\, F{R}\nn
									&=	\frac{1}{3}\frac{1}{\paren{4\pi}^2} \left\{ b_0^{h^\perp} Q_2[W_{h^\perp}^2]\int \df ^4x\sqrt{g}{R} \right\}
										  +\mO(Y,{R}^2)
}
with
\al{
Q_2[W_{h^\perp}^2]	
	&=	\frac{1}{\Gamma(2)}\int ^{\infty}_0\df z\, z\frac{\paren{\p_tF}R_\Lambda	+F\paren{\p_tR_\Lambda}}{(FP_\Lambda-V)^2}F\nn
	&=\frac{\Lambda^6F}{\paren{F\Lambda^2-V}^2}\left[ \frac{1}{6}\paren{\p_tF}-F\right].
			}
To summarize, we get the loop contributions of $h^\perp$ field:
\al{
\frac{1}{2}{\rm Tr}\left. \frac{\p_t{\mathcal R}_\Lambda}{\Gamma _\Lambda^{(1,1)}	
					+{\mathcal R}_\Lambda}\right|_{h^\perp h^\perp}
	&=	\frac{b_0^{h^\perp}}{2\paren{4\pi}^2} \frac{\Lambda^6}{F\Lambda^2-V}\left[ \frac{1}{6}(\p_t F) -F \right] \int \df ^4x\sqrt{g}\nn
	&\quad	+\frac{1}{\paren{4\pi}^2}\left\{ \frac{b_2^{h^\perp}}{2}\frac{ \Lambda^4}{F\Lambda^2-V}\left[ \frac{1}{2}\paren{\p_tF}-2F\right] 
						-\frac{b_0^{h^\perp}}{3}\frac{\Lambda^6F}{\paren{F\Lambda^2-V}^2}\left[ \frac{1}{6}\paren{\p_tF}-F\right]\right\}\nn
	&\qquad	\times\int \df ^4x\sqrt{g}{R} \nn					
	&\quad	+\frac{b_0^{h^\perp}}{\paren{4\pi}^2}	\frac{\Lambda^6}{\paren{F\Lambda^2-V}^2}\left[ \frac{1}{6}\paren{\p_tF}-F\right] \int \df ^4x\sqrt{g} Y
					+\mO(Y^2,{R}^2).
}
The first, second and third term are the contribution to $V$, $F$ and $Y=y\phi \ol \psi \psi$, respectively.
The correction to the Yukawa interaction corresponds to the diagram (I) in Fig.~\ref{yukawacorrections}.

\subsubsection{The loop contribution of the gravity field with spin 1} 
We evaluate the loop contribution of $\xi$ field:
\al{
\nonumber
\frac{1}{2}{\rm Tr}'\left. \frac{\p_t{\mathcal R}_\Lambda}{\Gamma _\Lambda^{(1,1)}		+{\mathcal R}_\Lambda}\right|_{\xi \xi}
		&=\frac{1}{2}{\rm Tr}'\frac{\paren{\p_tF}R_\Lambda+F\paren{\p_tR_\Lambda}}{F(P_\Lambda+\frac{2\alpha-1}{4}{R})-\alpha (V+Y)}\\
		&=\frac{1}{2}{\rm Tr}'	\frac{\paren{\p_tF}R_\Lambda+F\paren{\p_tR_\Lambda}}{FP_\Lambda-\alpha (V+Y)}
		-\frac{1}{8}{\rm Tr}'	\frac{\paren{\p_tF}R_\Lambda+F\paren{\p_tR_\Lambda}}{(FP_\Lambda-\alpha (V+Y))^2}F(2\alpha-1){R}
		+{\mathcal O}({R}^2).
}
We obtain
\al{
\frac{1}{2}{\rm Tr}[W_\xi^1]	&:= \frac{1}{2}{\rm Tr}	\frac{\paren{\p_tF}R_\Lambda+F\paren{\p_tR_\Lambda}}{FP_\Lambda-\alpha (V+Y)}\nn
						&=	\frac{1}{2}\frac{1}{\paren{4\pi}^2}	\left\{ b^\xi_0Q_2[W_\xi^1]\int \df^4x\sqrt{g}
							+b^\xi_2Q_1[W_\xi^1]\int \df^4x\sqrt{{g}}{R}\right\}	
								 +{\mathcal O}(R^2),
}
where
\al{
Q_2[W_\xi^1]	&=	\frac{1}{\Gamma(2)}\int ^{\infty}_0\df z\, z\frac{\paren{\p_tF}R_\Lambda	+F\paren{\p_tR_\Lambda}}{FP_\Lambda-\alpha (V+Y)}
			=	\frac{\Lambda^6}{F\Lambda^2-\alpha (V+Y)}\left[ \frac{1}{6}\paren{\p_tF}-F\right],\\
Q_1[W_\xi^1]	&=	\frac{1}{\Gamma(1)}\int ^{\infty}_0\df z\, \frac{\paren{\p_tF}R_\Lambda	+F\paren{\p_tR_\Lambda}}{FP_\Lambda-\alpha (V+Y)}
			=\frac{\Lambda^4}{F\Lambda^2-\alpha V}\left[ \frac{1}{2}\paren{\p_tF}-2F\right],
			}
and
\al{
-\frac{1}{8}{\rm Tr}[W_\xi^2]{R} &:= -\frac{1}{8}{\rm Tr}	\frac{\paren{\p_tF}R_\Lambda+F\paren{\p_tR_\Lambda}}{(FP_\Lambda-\alpha (V+Y))^2}F(2\alpha-1){R} \nn
		&=-\frac{1}{8}\frac{1}{\paren{4\pi}^2}\left\{ b^\xi_0Q_2[W_\xi^2]\int \df^4x\sqrt{g}{R}\right\}	+{\mathcal O}({R}^2),\nn
}
with
\al{
Q_2[W_\xi^2]	&=	\frac{1}{\Gamma(2)}	\int ^{\infty}_0\df z\, z\frac{\paren{\p_tF}R_\Lambda	+F\paren{\p_tR_\Lambda}}{(FP_\Lambda-\alpha (V+Y))^2}F(2\alpha-1)\nn
			&=\frac{\Lambda^6F(2\alpha-1)}{(F\Lambda^2-\alpha (V+Y))^2}\left[ \frac{1}{6}(\p_t F)-F\right].
}
We get the loop contributions of $\xi$ field:
\al{
\frac{1}{2}{\rm Tr}'\left. \frac{\p_t{\mathcal R}_\Lambda}{\Gamma _\Lambda^{(1,1)}		+{\mathcal R}_\Lambda}\right|_{\xi \xi}
	&=	\frac{b_0^\xi}{2\paren{4\pi}^2}\frac{\Lambda^6}{F\Lambda^2-\alpha (V+Y)}\bigg[ \frac{1}{6}\paren{\p_tF}-F\bigg] \int \df^4x\sqrt{g}\nn
&\quad+\frac{1}{\paren{4\pi}^2}\bigg\{ \frac{b_2^\xi}{2}\frac{\Lambda^4}{F\Lambda^2-\alpha V}\bigg[ \frac{1}{2}\paren{\p_tF}-2F\bigg]\nn
&\quad -\frac{b_0^\xi}{8}\frac{\Lambda^6F(2\alpha-1)}{(F\Lambda^2-\alpha (V+Y))^2}\bigg[ \frac{1}{6}(\p_t F)-F\bigg]
 					\bigg\}
					\int \df^4x\sqrt{g}{R}.
}
Obviously, when employing the de-Donder gauge $\alpha =0$, the terms with $\alpha$ vanish.
Thus, the correction exhibited as the diagram (II) in Fig.~\ref{yukawacorrections}  does not contribute to the beta functions in the de-Donder gauge.

\subsubsection{The loop contribution of the gravity fields with spin 0 and the scalar field}
Let us evaluate the spin 0 field loop contribution:
\al{
\frac{1}{2}{\rm Tr}''\left. \frac{\p_t{\mathcal R}_\Lambda}{\Gamma _\Lambda^{(1,1)}+{\mathcal R}_\Lambda}\right|_{\rm SS}.
}
We calculate the inverse matrix of $\Gamma _\Lambda^{(1,1)}+{\mathcal R}_\Lambda$, multiply as $(\Gamma _\Lambda^{(1,1)}+{\mathcal R}_\Lambda)^{-1}\,\p_t{\mathcal R}_\Lambda$, and take the de-Donder gauge to obtain
\al{
\frac{1}{2}{\rm Tr}''\left. \frac{\p_t{\mathcal R}_\Lambda}{\Gamma _\Lambda^{(1,1)}+{\mathcal R}_\Lambda}\right|_{\rm SS}
&=\frac{b_0^{\rm S}}{2\paren{4\pi}^2}Q_2[A] \int \df^4x\sqrt{g}
+\frac{1}{2\paren{4\pi}^2}\left\{ b_2^{\rm S} Q_1[A] + b_0^{\rm S} Q_2[B] \right\} \int \df^4x\sqrt{g}{R}
\nn
&\quad+\frac{b_0^{\rm S}}{2(4\pi)^2}Q_2[C] \int \df^4x \sqrt{g} Y +\mO({R}^2,Y^2),
}
where
\al{
\frac{b_0^{\rm S}}{2}\frac{1}{\paren{4\pi}^2}Q_2[A]
	&=	\frac{b_0^{\rm S}}{\paren{4\pi}^2}\frac{1}{\Gamma(2)}\int \df z\, z\,\theta\fn{\Lambda^2-z}\nn
&\quad\times \bigg[ -\frac{2}{2}\frac{F\Lambda^2(\Lambda^2\Psi+24\phi^2\Lambda^2F'^2\Psi'+F\Lambda^2\Sigma_1
	+(\Psi\Sigma_1+12\phi^2\Psi'^2))}{F\Lambda^2(\Psi \Sigma_1+12\Psi'^2\phi^2)}    \nn
&\phantom{\quad\times\bigg[}
	+\frac{1}{2}(\Lambda^2-z)\left( \frac{F\Lambda^2\Sigma_1+(\Psi\Sigma_1+12\Psi'^2\phi^2)}{F\Lambda^2(\Psi \Sigma_1+12\Psi'^2\phi^2)} \right)\p_tF\nn
&\phantom{\quad\times\bigg[}
	+\frac{24}{2}(\Lambda^2-z)\left( \frac{F\Lambda^2\phi^2\Psi'}{F\Lambda^2(\Psi \Sigma_1+12\Psi'^2\phi^2)} \right)	\p_tF' \bigg]\nn
&=\frac{b_0^{\rm S}\Lambda^4}{192\pi^2}\left[ -6-\frac{6(\Lambda^2\Psi+24\phi^2\Lambda^2F'\Psi'+F\Lambda^2\Sigma_1)}{\Delta} 
	+ \left( \frac{\Lambda^2\Sigma_1}{\Delta}+\frac{1}{F}\right)\p_tF+\left( \frac{24\phi^2\Lambda^2\Psi'}{\Delta}\right) \p_tF'\right],\\
\frac{b_2^{\rm S}}{2}\frac{1}{(4\pi )^2}Q_1[A]
	&=	\frac{b_2^{\rm S}}{\paren{4\pi}^2}\frac{1}{\Gamma(1)}\int \df z~\theta\fn{\Lambda^2-z}\nn
	&\quad\times \bigg[ -1-\frac{\Lambda^2\Psi+24\phi^2\Lambda^2F'\Psi'+F\Lambda^2\Sigma_1}{\Delta} \nn
	&\phantom{\quad\times \bigg[}
		+\frac{1}{2}(\Lambda^2-z)\left( \frac{\Sigma_1}{\Delta}+\frac{1}{F\Lambda^2}\right)\p_tF
		+\frac{24}{2}(\Lambda^2-z)\left( \frac{\phi^2\Psi'}{\Delta}\right) \p_tF'\bigg]\nn
	&=\frac{b_2^{\rm S}\Lambda^2}{16\pi^2}\left[ -1-\frac{\Lambda^2\Psi+24\phi^2\Lambda^2F'\Psi'+F\Lambda^2\Sigma_1}{\Delta}
		+ \frac{1}{4}\left( \frac{\Lambda^2\Sigma_1}{\Delta}+\frac{1}{F}\right)\p_tF
			+\frac{24}{4}\left( \frac{\Lambda^2\phi^2\Psi'}{\Delta}\right) \p_tF'\right],
}
and
\begin{align}
\frac{b_0^{\rm S}}{2}\frac{1}{\paren{4\pi}^2}Q_2[B]
	&=	\frac{b_0^{\rm S}}{(4\pi )^2}\frac{1}{\Gamma(2)}\int \df z\, z\theta\fn{\Lambda^2-z}\nn
	&\quad\times \bigg[ \frac{-\Delta^2 +4\phi^2(6\Lambda^4F'^2+\Psi'^2)\Delta
		-4\phi^2\Psi\Psi'\paren{7\Lambda^2F'-V'}(\Sigma_1-\Lambda^2)}{2\Lambda^2\Delta ^2} \nn
	&\phantom{\quad\times \bigg[}
		+\frac{-4\phi^2\Sigma_1\paren{7\Lambda^2F'-V'}(2\Psi V'-V\Psi ')
		+(-2\Lambda^4\Psi^2 -48\Lambda^4F'\phi^2\Psi \Psi'+24\Lambda^4F\phi^2\Psi'^2)\Sigma _2}{2\Lambda^2\Delta ^2} \nn
	&\phantom{\quad\times \bigg[}
		+(\Lambda^2-z)\left( \frac{\Delta^2 +4\phi^2 V'\Psi'\Delta 
		-24\Lambda^4F\phi^2\Psi'^2\Sigma _2-4\phi^2\Lambda^2F\Psi'\Sigma _1\paren{7\Lambda^2F'-V'}}{4F\Lambda^4\Delta ^2}\right)  \p_tF\nn
	&\phantom{\quad\times \bigg[}
		+4F\Lambda^2(\Lambda^2-z)\phi^2 \left(  \frac{-12\Psi'(7F'\Lambda^2-V')\phi^2
		 +(7F'\Lambda^2-V')\Psi \Sigma_1+12\Lambda^2\Psi\Psi'\Sigma_2}{4F\Lambda^4\Delta ^2}  \right)\p_tF'
		 \bigg]\nn
	&=	\frac{b_0^{\rm S}}{\paren{4\pi}^2} \bigg[
			\frac{\Lambda^2}{4}\frac{-\Delta^2 +4\phi^2(6\Lambda^4F'^2+\Psi'^2)\Delta
			-4\phi^2\Psi\Psi'\paren{7\Lambda^2F'-V'}(\Sigma_1-\Lambda^2)}{\Delta ^2} \nn
	&\phantom{=	\frac{b_0^{\rm S}}{\paren{4\pi}^2} \bigg[}
		+\frac{\Lambda^2}{4}
			\frac{-4\phi^2\Sigma_1\paren{7\Lambda^2F'-V'}(2\Psi V'-V\Psi ')}{\Delta ^2}\nn
	&\phantom{=	\frac{b_0^{\rm S}}{\paren{4\pi}^2} \bigg[}
		+\frac{\Lambda^2}{4}
				\frac{(-2\Lambda^4\Psi^2 -48\Lambda^4F'\phi^2\Psi \Psi'+24\Lambda^4F\phi^2\Psi'^2)\Sigma _2}{\Delta ^2}  \nn
	&\phantom{=	\frac{b_0^{\rm S}}{\paren{4\pi}^2} \bigg[}
		+\frac{\Lambda^2}{24}\left( \frac{\Delta^2 +4\phi^2 V'\Psi'\Delta 
		-24\Lambda^4F\phi^2\Psi'^2\Sigma _2-4\phi^2\Lambda^2F\Psi'\Sigma _1\paren{7\Lambda^2F'-V'}}{F\Delta ^2}\right)  \p_tF \nn
	&\phantom{=	\frac{b_0^{\rm S}}{\paren{4\pi}^2} \bigg[}
		+\frac{\Lambda^4}{6}\phi^2 \left(  \frac{-12\Psi'(7F'\Lambda^2-V')\phi^2 +(7F'\Lambda^2-V')\Psi \Sigma_1
		+12\Lambda^2\Psi\Psi'\Sigma_2}{\Delta ^2}  \right)\p_tF' 
		\bigg],
\end{align}
where $\Psi$, $\Sigma_1$, $\Sigma_2$ and $\Delta$ are given in (\ref{somedefinitions}).

The order $Y$ is given by
\al{\frac{b_0^{\rm S}}{2(4\pi)^2}Q_2[C]=
&\frac{\Lambda^6}{32\pi^2}\bigg[ 
		24\left( \xi_2 -\frac{\p_t \xi _2}{6}\right) I[1,1,0]
		-\left(\xi_0-\frac{\p_t\xi_0}{6}\right) I[2,0,0] \nn	
		&- \cred{12\,}C\left( \xi_0-\frac{\p_t\xi_0}{6}\right) I[2,1,0]
		-\cred{12\,}CI[1,2,0]
	\bigg],
}
where we have introduced
\al{
I[n_g,n_b,b_f]=\left. \frac{1}{\paren{F\Lambda^2-V}^{n_g}(\Lambda^2+2V'+4\phi^2V'')^{n_b}(\Lambda^2+y^2\phi^2)^{n_f}}\right| _{\phi=0},
}
$C=F'\Lambda^2-V'|_{\phi=0} =\xi_2 \Lambda^2 -\lambda_2$, and omitted the hat on the dimensionful coupling constants.
These terms include the corrections from the diagrams (III), (IV) and (V) in Fig.~\ref{yukawacorrections}. 
\subsection{Fermionic contributions}
We evaluate the fermionic loop contributions.
From (\ref{frgequation}), we have
\al{
&-{1\over2}\Tr\bigg[{\mc M}_\FF^{-1}\,{\p \mc R_\FF\over\p\Lambda}\bigg]\nn
		&-{1\over2}\Tr\bigg[{\mc M}_\FF^{-1}\,{\p \mc R_\FF\over\p\Lambda}\,{\mc M}_\FF^{-1}{\mc M}_\FB {\mc M}_\BB^{-1}{\mc M}_\BF\bigg]
		-{1\over2}\Tr\bigg[{\mc M}_\FF^{-1}{\mc M}_\FB {\mc M}_\BB^{-1}\,{\p \mc R_\BB\over\p\Lambda}\,{\mc M}_\BB^{-1}{\mc M}_\BF\bigg].
}
The first term contributes to the beta functions of $V$ and $F$ and the second and third terms contribute to the beta function of Yukawa coupling constant.

First we evaluate the first term:
\al{
\nonumber
-{1\over2}\Tr\bigg[{\mc M}_\FF^{-1}\,{\p \mc R_\FF\over\p t}\bigg]
=-{1\over2}\Tr\bigg[{\mc M}_\FF^{\rm phys}{}^{-1}\,{\p \mc R_\FF^{\rm phys}\over\p t}\bigg]
-{1\over2}\Tr\bigg[{\mc M}_\FF^{\rm ghost}{}^{-1}\,{\p \mc R_\FF^{\rm ghost}\over\p t}\bigg].
}

\subsubsection{The physical fermion contributions}
The physical part is calculated as
\al{
-{1\over2}\Tr\left[{\mc M}_\FF^{\rm phys}{}^{-1}\,{\p \mc R_\FF^{\rm phys}\over\p t}\right]
&=-{1\over2}\wt{\p\over\p t}\Tr \left[\ln \fn{{\mc M}_\FF^{\rm phys}} \right]
\nn
&= -{1\over2}\wt{\p\over\p t}\ln \Det
\pmat{
0	&	-{\Slash D}^\T 	-R_\Lambda\fn{{\Slash D}^\T}	-y\phi	\nn
{\Slash D}		+R_\Lambda\fn{{\Slash D}}	+y\phi	&	0
}\\
\nonumber
	&= -{\rm Tr}\left. \frac{\p_t{\mathcal R}_\Lambda}{\Gamma _\Lambda^{(1,1)}+{\mathcal R}_\Lambda}\right| _{{\ol \chi}\chi}\\
\nonumber
	&= -{\rm Tr} 
	\frac{\paren{\p_t{\mathcal R}^{(2,1)}_\Lambda}\left( -\frac{\sqrt{P_\Lambda\fn{p^2+{R\over4}}}}{\sqrt{p^2+\frac{R}{4}}}{\Slash D} +y\phi  \right)}
	{\left( \frac{\sqrt{P_\Lambda\fn{p^2+{R\over4}}}}{\sqrt{p^2+\frac{R}{4}}}{\Slash D} +y\phi \right) \left(-\frac{\sqrt{P_\Lambda\fn{p^2+{R\over4}}}}{\sqrt{p^2+\frac{R}{4}}}{\Slash D} +y\phi \right)}\\
\nonumber
	&=-{\rm Tr} 
	\frac{\left( \frac{\frac{1}{2}\p_tP_\Lambda\fn{p^2+{R\over4}}}{\sqrt{P_\Lambda\fn{p^2+{R\over4}}}\sqrt{p^2+\frac{R}{4}}}{\Slash D} \right)
	\left( -\frac{\sqrt{P_\Lambda\fn{p^2+{R\over4}}}}{\sqrt{p^2+\frac{R}{4}}}{\Slash D} +y\phi  \right)}
	{\left( \frac{P_\Lambda\fn{p^2+{R\over4}}}{p^2+\frac{R}{4}}\left( p^2+\frac{R}{4} \right) +y^2\phi^2 \right)}\\
\nonumber
	&= -{\rm Tr} \frac{\frac{1}{2}\p_t\,P_\Lambda\fn{p^2+{R\over4}}}{P_\Lambda\fn{p^2+{R\over4}}+y^2\phi^2}
	=	-\frac{1}{2}{\rm Tr}\frac{\p_t R_\Lambda}{P_\Lambda\fn{p^2+{R\over4}}+y^2\phi^2}
	=:-\frac{1}{2}{\rm Tr}[W_{\rm f}]\\
	&=-\frac{1}{2}\frac{1}{\paren{4\pi}^2} 
		\left\{ b_0^{\rm f}\,Q_2[W_{\rm f}]\int \df^dx\sqrt{g}
							+b_2^{\rm f}\,Q_1[W_{\rm f}] \int \df^dx\sqrt{g}{R} \right\}
								+{\mathcal O}({R}^2),
}
where we have used $-{\Slash D}^2=p^2 +{R}/4$ and 
\al{
\nonumber
Q_2[W_{\rm f}]&=\frac{1}{\Gamma(2)}\int ^{\infty}_0\df z\, z \frac{\p _tP_\Lambda}{P_\Lambda+y^2\phi^2}
\nonumber
				=\int ^{\Lambda^2}_0\df z\, z\frac{-2\Lambda^2}{\Lambda^2+y^2\phi^2}
				=\frac{-\Lambda^6}{\Lambda^2+y^2\phi^2} ,\\
\nonumber
Q_1[W_{\rm f}]&=\frac{1}{\Gamma(1)}\int ^{\infty}_0 \df z \frac{\p_tP_\Lambda}{P_\Lambda+y^2\phi^2} 
			 	=\int ^{\Lambda^2}_0 \df z \frac{-2\Lambda^2}{\Lambda^2+y^2\phi^2}
\nonumber
				=\frac{-2\Lambda^4}{\Lambda^2+y^2\phi^2}.
}
We obtain
\al{
-{1\over2}\Tr\bigg[{\mc M}_\FF^{\rm phys}{}^{-1}\,{\p \mc R_\FF^{\rm phys}\over\p t}\bigg]
&=-\frac{b_0^{\rm f}}{2\paren{4\pi}^2}\frac{-\Lambda^6}{\Lambda^2+y^2\phi^2} \int \df^dx\sqrt{g}\nn
&\quad -\frac{b_2^{\rm f}}{2\paren{4\pi}^2}\frac{-2\Lambda^4}{\Lambda^2+y^2\phi^2} \int \df^dx\sqrt{g}{R} 
+\mO ({R}^2).
}

\subsubsection{The ghost fields contributions}
We evaluate the ghost field contributions:
\al{
\nonumber
-{1\over2}\Tr\bigg[{\mc M}_\FF^{\rm ghost}{}^{-1}\,{\p \mc R^{\rm ghost}_\FF\over\p t}\bigg]
\nonumber
&=-{1\over2}\wt{\p\over\p t}\Tr \left[\ln \fn{{\mc M}_\FF^{\rm ghost}} \right]
=-{1\over2}\wt{\p\over\p t}\ln \fn{\Det \left[ {\mc M}_\FF^{\rm ghost} \right]}\\
\nonumber
&=
-{\rm Tr}\left. \frac{\p_t{\mathcal R}_\Lambda}{\Gamma _\Lambda^{(1,1)}+{\mathcal R}_\Lambda}\right|_{{\bar C}^\perp C^\perp}
					-{\rm Tr}'\left. \frac{\p_t{\mathcal R}_\Lambda}{\Gamma _\Lambda^{(1,1)}
						+{\mathcal R}_\Lambda}\right|_{\bar C C}.
}
For spin 1 ghost field, we have
\al{
-{\rm Tr}\left. \frac{\p_t{\mathcal R}_\Lambda}{\Gamma _\Lambda^{(1,1)}+{\mathcal R}_\Lambda}\right|_{C^\perp C^\perp}	
	=-{\rm Tr}\frac{\p_tR_\Lambda}{P_\Lambda-\frac{R}{4}}
	=-{\rm Tr}\frac{\p_tR_\Lambda}{P_\Lambda}-{\rm Tr}\frac{\p_tR_\Lambda}{P_\Lambda^2}\frac{R}{4}+{\mathcal O}({R}^2),
}
with
\al{
-{\rm Tr}\frac{\p_tR_\Lambda}{P_\Lambda}&=-\frac{1}{\paren{4\pi}^2}\left\{ b^{C^\perp}_0Q_2\left[ \frac{\p_tR_\Lambda}{P_\Lambda}\right]\int \df^4x\sqrt{g}
	+b^{C^\perp}_2Q_1\left[ \frac{\p_tR_\Lambda}{P_\Lambda}\right]\int \df^4x\sqrt{g}{R}\right\}
		 +{\mathcal O}({R}^2),\\
Q_2\left[ \frac{\p_tR_\Lambda}{P_\Lambda}\right]	&=	\frac{1}{\Gamma(2)}\int ^{\infty}_0\df z \, z\frac{\p_tR_\Lambda}{P_\Lambda}
=\int ^{\Lambda^2}_0\df z \, z\frac{-2\Lambda^2}{\Lambda^2}=-\Lambda^4,\\
Q_1\left[ \frac{\p_tR_\Lambda}{P_\Lambda}\right]	&=	\frac{1}{\Gamma(1)}\int ^{\infty}_0\df z \, \frac{\p_tR_\Lambda}{P_\Lambda}
=\int ^{\Lambda^2}_0\df z \, \frac{-2\Lambda^2}{\Lambda^2}=-2\Lambda^2,
}
and
\al{
-{\rm Tr}\frac{\p_tR_\Lambda}{P_\Lambda^2}\frac{R}{4}		&=	-\frac{1}{\paren{4\pi}^2}
	\left\{ \frac{b^{C^\perp}_0}{4}Q_2\left[ \frac{\p_tR_\Lambda}{P_\Lambda^2}\right] \int \df^4x\sqrt{g}{R}\right\} 
	+{\mathcal O}({R}^2),\\
Q_2\left[ \frac{\p_tR_\Lambda}{P_\Lambda^2}\right]	&=	\frac{1}{\Gamma(2)}\int ^{\infty}_0\df z\, z\frac{\p_tR_\Lambda}{P_\Lambda^2}
	=\int ^{\Lambda^2}_0\df z~z\frac{-2\Lambda^2}{\Lambda^4}	=-\Lambda^2.
}
We obtain the contributions of the ghost field with spin 1:
\al{
\nonumber
-{\rm Tr}\left. \frac{\p_t{\mathcal R}_\Lambda}{\Gamma _\Lambda^{(1,1)}+{\mathcal R}_\Lambda}\right|_{C^\perp C^\perp}	
	&=-\frac{b_0^{C^\perp}(-\Lambda^4)}{\paren{4\pi}^2} \int \df^4x\sqrt{g}\\
 	&-\frac{1}{\paren{4\pi}^2}\left\{ b_2^{C^\perp}(-2\Lambda^2)	+ \frac{b_0^{C^\perp}}{4} \paren{-\Lambda^2} \right\} \int \df^4x\sqrt{g}{R}.
}

For spin 0 ghost field, we have
\al{
-{\rm Tr}'\left. \frac{\p_t{\mathcal R}_\Lambda}{\Gamma _\Lambda^{(1,1)}	+{\mathcal R}_\Lambda}\right|_{{\bar C}{C}}
	&=-{\rm Tr}'\frac{(2-\frac{1+\beta}{2})\p_tR_\Lambda}{(2-\frac{1+\beta}{2})P_\Lambda-\frac{R}{2}}
	=-{\rm Tr}\frac{\p_tR_\Lambda}{P_\Lambda}
			-{\rm Tr}\frac{\p_tR_\Lambda}{P_\Lambda^2}\frac{R}{3-\beta}+{\mathcal O}({R}^2),
}
with
\al{
-{\rm Tr}\frac{\p_tR_\Lambda}{P_\Lambda}&=-\frac{1}{\paren{4\pi}^2}\left\{ b^{C}_0 Q_2\left[ \frac{\p_tR_\Lambda}{P_\Lambda}\right]\int \df^4x\sqrt{g}
		+b^{C}_2Q_1\left[ \frac{\p_tR_\Lambda}{P_\Lambda}\right]\int \df^4x\sqrt{g}{R}\right\}
			 +{\mathcal O}({R}^2),\\
Q_2\left[ \frac{\p_tR_\Lambda}{P_\Lambda}\right]		&=\frac{1}{\Gamma(2)}\int ^{\infty}_0\df z~z\frac{\p_tR_\Lambda}{P_\Lambda}
								=\int ^{\Lambda^2}_0\df z~z\frac{-2\Lambda^2}{\Lambda^2}=-\Lambda^4,\\
Q_1\left[ \frac{\p_tR_\Lambda}{P_\Lambda}\right]		&=\frac{1}{\Gamma(1)}\int ^{\infty}_0\df z~\frac{\p_tR_\Lambda}{P_\Lambda}
								=\int ^{\Lambda^2}_0\df z~\frac{-2\Lambda^2}{\Lambda^2}=-2\Lambda^2,
}
and
\al{
-{\rm Tr}\frac{\p_tR_\Lambda}{P_\Lambda^2}\frac{R}{3-\beta}
	&=-\frac{1}{\paren{4\pi}^2}
		\left\{\frac{b^{C}_0}{3-\beta} Q_2\left[ \frac{\p_tR_\Lambda}{P_\Lambda^2}\right] \int \df^4x\sqrt{g}{R}\right\}
	 		+{\mathcal O}({R}^2),\\
Q_2\left[ \frac{\p_tR_\Lambda}{P_\Lambda^2}\right]	&=\frac{1}{\Gamma(2)}\int ^{\infty}_0\df z~z\frac{\p_tR_\Lambda}{P_\Lambda^2}
								=\int ^{\Lambda^2}_0\df z~z\frac{-2\Lambda^2}{\Lambda^4}=-\Lambda^2.
}
We obtain the ghost field contributions:
\al{
\nonumber
-{\rm Tr}'\left. \frac{\p_t{\mathcal R}_\Lambda}{\Gamma _\Lambda^{(1,1)}	+{\mathcal R}_\Lambda}\right|_{{\bar C}{C}}
&=-\frac{b_0^C (-\Lambda^4)}{\paren{4\pi}^2} \int \df^4x\sqrt{g} \\
&\quad
	-\frac{1}{\paren{4\pi}^2} \left\{ b_2^C (-2\Lambda^2) + \frac{b_0^C}{3-\beta} \paren{-\Lambda^2} \right\} \int \df^4x\sqrt{g}{R}.
}

\subsection{Contribution from both fermion and boson}
We evaluate the terms
\al{
		-{1\over 2}\Tr\bigg[{\mc M}_\FF^{-1}\,{\p \mc R_\FF\over\p\Lambda}\,{\mc M}_\FF^{-1}{\mc M}_\FB {\mc M}_\BB^{-1}{\mc M}_\BF\bigg]
		-{1\over 2}\Tr\bigg[{\mc M}_\FF^{-1}{\mc M}_\FB {\mc M}_\BB^{-1}\,{\p \mc R_\BB\over\p\Lambda}\,{\mc M}_\BB^{-1}{\mc M}_\BF\bigg],
}
which contribute to the beta function of the Yukawa coupling constant.
We obtain the corrections (VI)--(XII) described by Fig.~\ref{yukawacorrections}.
Note that since the diagram (VII) vanishes when employing the de-Donder gauge $\alpha=0$ and $\beta=1$, we ignore it here.

First, we evaluate the diagram (VI):
\al{
\nonumber
&\Tr \bigg[ \frac{y\phi (-\p_t R_\Lambda)}{(P_\Lambda+y^2\phi^2)^2(P_\Lambda+M_\phi^2)} 
+ \frac{y\phi (-\p_t R_\Lambda)}{(P_\Lambda+y^2\phi^2)(P_\Lambda+M_\phi^2)^2} 
\bigg](y\psi)(y{\ol \psi}) \nn
&\quad =-\frac{y^2\Lambda^6}{16\pi^2}\left(  I[0,1,2]+I[0,2,1]\right)\int \df^4x \sqrt{{g}}\, Y,
}
Second, we evaluate the diagrams (VIII) and (IX) in Fig~\ref{yukawacorrections}:
\al{
\nonumber
&\Tr \Bigg[ 
\bigg\{
\frac{y\phi (-\p_t R_\Lambda)}{(P_\Lambda+y^2\phi^2)^2(FP_\Lambda -V)}
+ \frac{y\phi  ( (\p_t F) R_\Lambda+F\p_t R_\Lambda)}{(P_\Lambda+y^2\phi^2)(FP_\Lambda-V)^2}
\bigg\}
\nn
& \quad \times
\pmat{
\displaystyle\frac{3}{16}(\gamma^\mu \psi) \p_\mu &\displaystyle\frac{3}{16} (\gamma^\mu \psi) \p_\mu
}
\pmat{
-\frac{4}{3} & 4\\
4 & -12
}
\pmat{
-\displaystyle\frac{3}{16}\paren{\ola{\p_\nu}}\paren{\ol\psi\gamma^\nu}\\[10pt]
-\displaystyle\frac{3}{16}\paren{\ola{\p_\nu}}\paren{\ol\psi\gamma^\nu}
}
\Bigg]
\nn
&\quad =-\frac{\Lambda^8}{128\pi^2}\Bigg[ I[1,0,2]+ \left( \xi_0 -\frac{\p_t\xi_0}{8} \right) I[2,0,1]\Bigg] \int \df^4x \sqrt{{g}}\, Y,
}
where we employed the de-Donder gauge $\alpha=0$ and $\beta =1$.
Third, we evaluate the diagrams (X) and (XI) in Fig~\ref{yukawacorrections}:
\al{\nonumber
&\Tr \Bigg[
\bigg\{
-\frac{\p_t R_\Lambda\,\left(\sqrt{P_\Lambda} -\frac{P_\Lambda+y^2\phi^2}{2\sqrt{P_\Lambda}}\right)\frac{\Slash \p}{z}}{(P_\Lambda +y^2\phi^2)^2(FP_\Lambda-V)}
-\frac{ (\p_t F R_\Lambda +F \p_t R_\Lambda ) \sqrt{P_\Lambda} {\Slash \p}   }{(P_\Lambda +y^2\phi^2)(FP_\Lambda-V)^2}
\bigg\}
\nn
&\quad \times
\begin{pmatrix}
\displaystyle \frac{3}{16} (\gamma^\mu \psi) \p_\mu	&	\displaystyle\frac{y}{2}\phi \psi +\frac{3}{16} (\gamma^\mu \psi) \p_\mu
\end{pmatrix}
\begin{pmatrix}
-\frac{4}{3}	&	4 \\
4	&	-12
\end{pmatrix} 
\begin{pmatrix}
-\displaystyle\frac{3}{16}\paren{\ola{\p_\nu}}\paren{\ol\psi\gamma^\nu}	\\[10pt]
\displaystyle\frac{y}{2}\phi {\bar \psi} -\displaystyle\frac{3}{16}\paren{\ola{\p_\nu}}\paren{\ol\psi\gamma^\nu}
\end{pmatrix}
\Bigg]
 \nn
&\xrightarrow[{\rm Projecting~to}~Y]{}  \frac{3\Lambda^8}{40\pi^2} \left[ I[1,0,2] + \left( \xi_0 -\frac{\p_t \xi_0}{7} \right) I[2,0,1] -\frac{1}{2\Lambda^2} I[1,0,1] \right] \int \df^4x \sqrt{{g}}\, Y.
}
Finally, we evaluate the diagrams (XII) and (XIII):
\al{
&\Tr\Bigg[
\bigg\{
 - \frac{\p_t R_\Lambda\, \left(\sqrt{P_\Lambda} -\frac{P_\Lambda+y^2\phi^2}{2\sqrt{P_\Lambda}}  \right)
(F' P_\Lambda -V')\phi\, {\Slash \p}}{(P_\Lambda +y^2\phi^2)^2(FP_\Lambda-V)\Sigma_1}
+\frac{ \sqrt{P_\Lambda}(\p_tR_\Lambda (F'P_\Lambda -V') )  \phi\, {\Slash \p}}{(P_\Lambda +y^2\phi^2)(FP_\Lambda-V)\Sigma_1^2}
\nn
&\quad -
\frac{ \left\{ [ (\p_t F')  (FP_\Lambda -V) + (\p_t F )(F'P_\Lambda -V') ]R_\Lambda  +(F V' -V F') \p_t R_\Lambda  \right\} \sqrt{P_\Lambda}  \phi\, {\Slash \p}}{(P_\Lambda +y^2\phi^2)(FP_\Lambda-V)^2\Sigma_1}
\bigg\}
\nn
&\quad \times 
\begin{pmatrix}
\displaystyle \frac{3}{16}(\gamma^\mu \psi)\p_\mu	&	\displaystyle \frac{3}{16}(\gamma^\mu \psi	) \p_\mu &	y\psi
\end{pmatrix} 
\pmat{
\#&\#&4\\
\#&\#&-12\\
4&-12& \# \\ 
}
\begin{pmatrix}
-\displaystyle\frac{3}{16}\paren{\ola{\p_\nu}}\paren{\ol\psi\gamma^\nu} \\[10pt]
-\displaystyle\frac{3}{16}\paren{\ola{\p_\nu}}\paren{\ol\psi\gamma^\nu} \\[10pt]
y\ol \psi 
\end{pmatrix}
\Bigg]
\nn
&\quad \xrightarrow[{\rm Projecting~to}~Y]{} 
-\frac{3\Lambda^8}{20\pi^2}\Bigg[
	-\left( \xi_0 -\frac{\p_t\xi_0}{7} \right) CI[2,1,1]	 +\left( \xi_2 -\frac{\p_t \xi_2}{7} \right)I[1,1,1]  \nn
	&\phantom{\quad \xrightarrow[{\rm Projecting~to}~Y]{} -}
		-C\left( I[1,2,1] + I[1,1,2]	- \frac{1}{2\Lambda^2} I[1,1,1] \right)
		\Bigg]\int \df^4x\sqrt{g} Y.
}

The beta functions for $V$, $F$ and $y$ in our truncated effective action are given as the coefficients of $\int d^4x \sqrt{g}$, $\int d^4x \sqrt{g}R$, and $\int d^4x\, \phi \ol \psi \psi=\int d^4x\, Y/y$, respectively.
Then we obtain the beta functions \eqref{betapotential}, \eqref{betanonminimal} and \eqref{betayukawa}.

\bibliography{refs}

\bibliographystyle{TitleAndArxiv}

\end{document}